\documentclass[sigconf]{acmart}

\copyrightyear{2021}
\acmYear{2021}
\setcopyright{acmcopyright}\acmConference[CCS '21]{Proceedings of the 2021 ACM
SIGSAC Conference on Computer and Communications Security}{November 15--19,
2021}{Virtual Event, Republic of Korea}
\acmBooktitle{Proceedings of the 2021 ACM SIGSAC Conference on Computer and
Communications Security (CCS '21), November 15--19, 2021, Virtual Event, Republic
of Korea}
\acmPrice{15.00}
\acmDOI{10.1145/3460120.3484585}
\acmISBN{978-1-4503-8454-4/21/11}

\AtBeginDocument{%
  \providecommand\BibTeX{{%
    \normalfont B\kern-0.5em{\scshape i\kern-0.25em b}\kern-0.8em\TeX}}}

\usepackage{amsmath,amsfonts}
\usepackage{textcomp}
\usepackage{xcolor}

\graphicspath{{./Figures/}}
\usepackage{subfigure}
\usepackage{graphicx}

\newcommand{\etal}{{\textit{et al. }}}
\newcommand{\NM}{\textsf{Whisper}}
\newcommand{\NMM}{Whisper}

\usepackage{amssymb}
\usepackage{mathrsfs}
\usepackage{bm}

\usepackage{multirow}
\newcommand{\tabincell}[2]{\begin{tabular}{@{}#1@{}}#2\end{tabular}}

\begin{document}
\fancyhead{} 

\title{Realtime Robust Malicious Traffic Detection\\via Frequency Domain Analysis}

\author{Chuanpu Fu$^{1}$, Qi Li$^{2,3}$, Meng Shen$^{4}$, and Ke Xu$^{1,3,5}$} 
\affiliation{
$^{1}$Department of Computer Science and Technology, Tsinghua University, Beijing, China\\
$^{2}$Institute for Network Sciences and Cyberspace, Tsinghua University, Beijing, China\\
$^{3}$Beijing National Research Center for Information Science and Technology (BNRist), Tsinghua  University, Beijing, China\\
$^{4}$School of Cyberspace Science and Technology, Beijing Institute of Technology, Beijing, China\\
$^{5}$Peng Cheng Laboratory, China\\
\{fcp20@mails., qli01@, xuke@\}tsinghua.edu.cn, shenmeng@bit.edu.cn
\country{} 
}

\begin{abstract}
Machine learning (ML) based malicious traffic detection is an emerging security paradigm, particularly for zero-day attack detection, which is complementary to existing rule based detection. However, the existing ML based detection achieves low detection accuracy and low throughput incurred by inefficient traffic features extraction. Thus, they cannot detect attacks in realtime, especially in high throughput networks. Particularly, these detection systems similar to the existing rule based detection can be easily evaded by sophisticated attacks. To this end, we propose \NM, a realtime ML based malicious traffic detection system that achieves both high accuracy and high throughput by utilizing frequency domain features. It utilizes sequential information represented by the frequency domain features to achieve bounded information loss, which ensures high detection accuracy, and meanwhile constrains the scale of features to achieve high detection throughput. In particular, attackers cannot easily interfere with the frequency domain features and thus \NM\ is robust against various evasion attacks. Our experiments with 42 types of attacks demonstrate that, compared with the state-of-the-art systems, \NM\ can accurately detect various sophisticated and stealthy attacks, achieving at most 18.36\% improvement of AUC, while achieving two orders of magnitude throughput. Even under various evasion attacks, \NM\ is still able to maintain around 90\% detection accuracy.  
\end{abstract}



\settopmatter{printacmref=false}
\maketitle

\newcommand{\sizeref}{\fontsize{8}{8}\selectfont}
\textbf{\sizeref ACM Reference Format:}\\
{\sizeref Chuanpu Fu, Qi Li, Meng Shen, and Ke Xu. 2021. Realtime Robust Malicious Traffic Detection via Frequency Domain Analysis. In \textit{Proceedings of the 2021 ACM SIGSAC Conference on Computer and Communications Security (CCS ’21), November 15–19, 2021, Virtual Event, Republic of Korea.} ACM, New York, NY, USA, 16 pages. \url{https://doi.org/10.1145/3460120.3484585}}

\section{Introduction} \label{section:introduction}
Traditional malicious traffic detection identifies malicious traffic by analyzing the features of traffic according to preconfigured rules, which aims to protect legitimate Internet users from network attacks~\cite{RAID15-Haetae, CCS12-Kargus}. However, the rule-base detection is unable to detect zero-day attacks~\cite{C&S09-survey, ACM-CS-16-survey, IEEE-CST-16-survey, INFO20-ZeroWall} though it can achieve high detection accuracy and detection throughput in high bandwidth networks, e.g., in Internet backbone networks. 

As a promising security paradigm, machine learning based malicious traffic detection has been developed, particularly as a complement of the traditional fixed rule based methods (i.e., signature based NIDS)~\cite{CCS18-vNIDS, CCS12-Kargus, USEC12-Chimera, RAID15-Haetae}. Table~\ref{table:packet-flow} summarizes and compares rule based and typical machine learning based detection methods. Compared with rule based methods, machine learning based methods can effectively identify zero-day malicious traffic~\cite{ACM-CS-16-survey, C&S09-survey}. Unfortunately, due to the processing overhead of machine learning algorithms, existing detection methods achieve low detection throughput and are unable to process high-rate traffic. As a result, most of these methods can only be deployed offline~\cite{USEC12-Botnet-Traffic, ACSAC12-Disclose, CCS19-Lifelong, USEC16-Variants, USEC15-WebWitness, NDSS14-Nazca} so that they cannot realize realtime detection, particularly in high performance networks (e.g., in 10 Gigabit networks)~\cite{NDSS18-Kitsune, Conext20-Qian, TIFS18-Qi}. 

Meanwhile, attackers can easily interfere with and evade these methods by injecting noises, e.g., packets generated by benign applications, into attack traffic. Packet-level detection~\cite{NDSS18-Kitsune, PRE18-rnn, RAID04-payload} that analyzes per-packet feature sequences is unable to achieve robust detection. Actually, even in the absence of the evasion attacks, the packet-level detection is unable to detect sophisticated zero-day attacks. 
Traditional flow-level methods~\cite{USEC16-Variants, NDSS14-Nazca, USEC15-WebWitness, TIFS18-Qi} detecting attacks by analyzing flow-level statistics incur significant detection latency.
Moreover, evasion attacks can easily bypass the traditional flow-level detection that uses coarse-grained flow-level statistics~\cite{S&P10-Outside, CCS04-Sommer}. Thus, realtime robust machine learning based detection that is ready for real deployment is still missing.

\newcommand{\bst}[1]{\color[rgb]{0.15, 0.616, 0.15}#1}
\newcommand{\wor}[1]{\color[rgb]{0.753,0,0}#1}

\newcommand{\cmark}{\bm{$\checkmark$}}
\newcommand{\xmark}{\bm{$\times$}}

\newcommand{\y}{\bst{\cmark}}
\newcommand{\n}{\wor{\xmark}}

\renewcommand{\arraystretch}{1.22}
\begin{table*}[t]
    \small
    \caption{Comparing the Existing Malicious Traffic Detection Methods}
    \vspace{-4mm}
    \begin{center}
        \begin{tabular}{@{}c|c|c|ccccccc@{}}
        \toprule
        \multicolumn{2}{c|}{\textbf{\tabincell{c}{Category of \\Detection Systems}}} & \textbf{\tabincell{c}{Feature Extraction Methods}} & \textbf{\tabincell{c}{Zero-Day\\Detection}} & \textbf{\tabincell{c}{High\\Accuracy}} & \textbf{\tabincell{c}{Robust\\Detection}} & \textbf{\tabincell{c}{Realtime\\Detection}} & \textbf{\tabincell{c}{High\\Throughput}} & \textbf{\tabincell{c}{Task\\Agnostic}} \\
        \midrule
        \multicolumn{2}{c|}{Rule based} & Preconfigured fix rules~\cite{CCS18-vNIDS, CCS12-Kargus, USEC12-Chimera} & \n & \y & \n & \y & \y & \n \\
        \midrule
        \multirow{6}{*}{ML based} & \multirow{3}{*}{Packet-level} & \tabincell{c}{Packet header fields~\cite{PRE18-rnn}} & \y & \y & \n & \y & \n & \y \\
        \cline{3-9}
        & & \tabincell{c}{Context statistics \cite{NDSS18-Kitsune}} & \y & \y & \n & \y & \n & \y \\
        \cline{3-9}
        & & \tabincell{c}{Payload statistics \cite{RAID04-payload}} & \y & \y & \n & \n & \n & \y \\
        \cline{2-9}
        & \multirow{3}{*}{Flow-level} & \tabincell{c}{Flow-level statistics~\cite{RAID19-SCADA, TIFS18-Qi, ACSAC12-Disclose}} & \y & \n & \n & \n & \y & \n \\
        \cline{3-9}
        & & \tabincell{c}{Application usage statistics~\cite{USEC16-Variants, USEC15-WebWitness, NDSS14-Nazca}} & \y & \y & {\n}$^{\rm 1}$ & \n & \n & \n \\
        \cline{3-9}
        & & \textbf{\tabincell{c}{Frequency domain features, \NM}} & \y & \y & \y & \y & \y & \y \\
        \bottomrule
        \multicolumn{7}{l}{\footnotesize $^{\rm 1}$ Bartos~\etal~\cite{USEC16-Variants} only considered evasion strategies for malicious Web traffic.}
        \end{tabular}
    \label{table:packet-flow}
    \end{center}
    \vspace{-3mm}
\end{table*}

In this paper, we develop \NM\ that aims to realize realtime robust malicious traffic detection by utilizing machine learning algorithms. \NM\ effectively extracts and analyzes the sequential information of network traffic by frequency domain analysis~\cite{SCI-COMP}, which extracts traffic features with low information loss. Especially, the frequency domain features of traffic can efficiently represent various packet ordering patterns of traffic with low feature redundancy. Frequency domain feature analysis with low information loss enables accurate and robust detection, while low feature redundancy ensures high throughput traffic detection. In particular, since the frequency domain features represent fine-grained sequential information of the packet sequences, which are not disturbed by the injected noise packets, \NM\ can achieve robust detection. However, it is non-trivial to extract and analyze the frequency domain features from traffic because of the large-scale, complicated, and dynamic patterns of traffic~\cite{S&P10-Outside, CCS04-Sommer}. 

To effectively perform frequency domain traffic feature analysis, we develop a three-step frequency domain feature extraction. First, we encode per-packet feature sequences as vectors, which reduces the data scale and the overhead of subsequent processing. Second, we segment the encoded vectors and perform Discrete Fourier Transformation (DFT)~\cite{SCI-COMP} on each frame, which aims to extract the sequential information of traffic. It allows statistical machine learning algorithms to easily learn the patterns. Third, we perform logarithmic transformation on the modulus of the frequency domain representation produced by DFT, which prevents float point overflows incurred by the numerical instability issue~\cite{DeepLearning} during the training of machine learning.

Furthermore, we propose an automatic parameter selection module to select the encoding vector for efficient packet feature encoding. To achieve this, we formulate the per-packet feature encoding as a constrained optimization problem to minimize mutual interference of the per-packet features during frequency domain feature analysis. We transform the original problem into an equivalent SMT problem and solve the problem by an SMT solver. It ensures the detection accuracy by choosing vectors, while effectively reducing manual efforts of selecting encoding vectors. We utilize statistical machine learning to cluster the patterns according to the frequency domain features. Due to the rich feature presentation and lightweight machine learning, \NM\ finally realizes realtime detection of malicious traffic in high throughput networks. 

We theoretically prove that \NM\ is more efficient than packet-level and traditional flow-level detection methods. We conduct a theoretical analysis to prove that the frequency domain features ensure bounded information loss, which lays the foundation for robust detection of \NM. We develop a \textit{traffic feature differential entropy model}, a theoretical framework to measure information loss of feature extraction from traffic. First, we prove the information loss in processing packet sequences in the existing flow-level methods, which further demonstrates that it cannot accurately extract features. Second, we prove that \NM\ maintains the information loss in the flow-level methods and validate that the frequency domain features are more efficient. Third, we prove that \NM\ effectively reduces feature redundancy by the decrease in the data scale of features. 

We prototype \NM\ with Intel’s Data Plane Development Kit (DPDK)~\cite{DPDK}. To extensively evaluate the performance of the \NM\ prototype, we replay 42 malicious traffic datasets with the high throughput backbone network traffic. Besides the typical attacks, we collect and replay 36 new malicious traffic datasets including: (i) more stealthy attacks, e.g., low-rate TCP DoS attacks~\cite{TON06-LRTCPDOS, NDSS10-LRTCPDOS, NDSS18-TCPWN} and stealthy network scanning~\cite{USEC19-SCAN}; (ii) complicated multi-stage attacks, e.g., TCP side-channel attacks~\cite{CCS20-MySC, USEC16-ACKSC, TON18-ACKSC} and TLS padding oracle attacks~\cite{EUROCRYPT02-TLS}; (iii) evasion attacks, i.e., attackers inject different types of noise packets (i.e., packets generated by various benign applications) in attack traffic to evade detection. According to our experimental results, we validate that \NM\ can detect the different types of attacks with AUC ranging between 0.891 and 0.999 while achieving 1,310,000 PPS, i.e., two orders of magnitude throughput more than the state-of-the-art methods. Particularly, \NM\ can detect various evasion attacks with 35\% improvement of AUC over the state-of-the-art methods. Furthermore, \NM\ achieves realtime detection with bounded 0.06 second detection latency in high throughput networks.

\begin{figure*}[t]
	\begin{center}
	\hspace{-12mm}
	\includegraphics[width=0.93\textwidth]{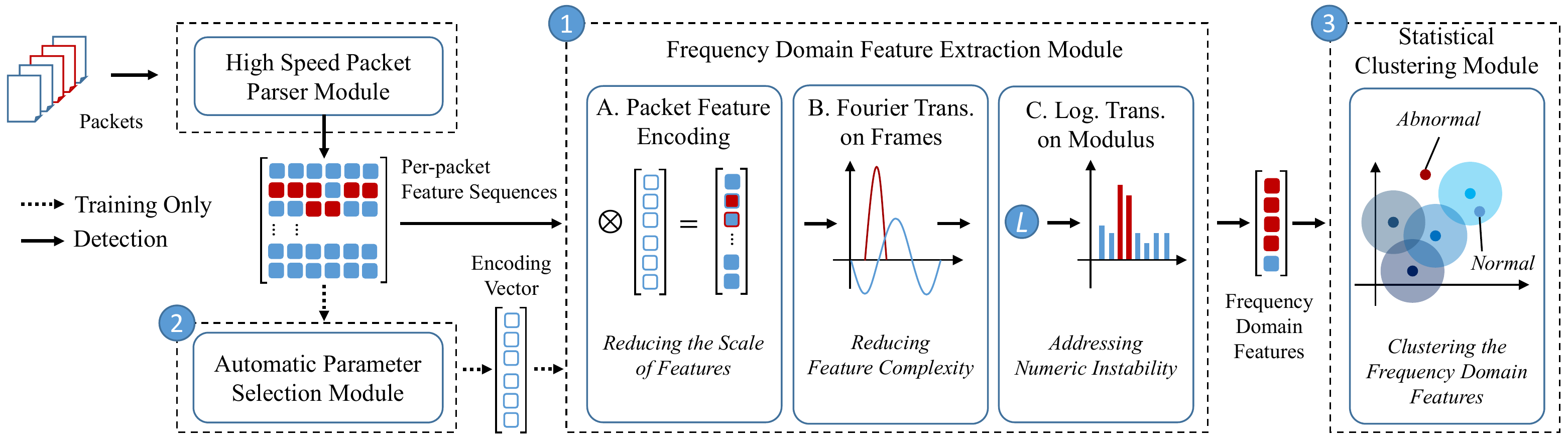}
	\vspace{-2mm}
	\caption{High-level design of \NM.}
	\label{diagram:high-level}
	\end{center}
	\vspace{-3mm}
\end{figure*}

In summary, the contributions of our paper are five-fold:
\begin{itemize}
\item We present \NM, a novel malicious traffic detection system by utilizing frequency domain analysis, which is the first system built upon machine learning achieving realtime and robust detection in high throughput networks. 
\item We perform frequency domain feature analysis to extract the sequential information of traffic, which lays the foundation for the detection accuracy, robustness, and high throughput of \NM.  
\item We develop automatic encoding vector selection for \NM\ to reduce manual efforts for parameter selection, which ensures the detection accuracy while avoiding manual parameter setting. 
\item We develop a theoretical analysis framework to prove the properties of \NM. 
\item We prototype \NM\ with Intel DPDK and use the experiments with different types of replayed attack traffic to validate the performance of \NM. 
\end{itemize}

The rest of the paper is organized as follows: Section~\ref{section:problem} introduces the threat model and the design goals of \NM. Section~\ref{section:overview} presents the high-level design of \NM. In Section~\ref{section:detail}, we present the design details of \NM. In Section~\ref{section:analysis}, we conduct a theoretical analysis. In Section~\ref{section:evaluation}, we experimentally evaluate the performances of \NM. Section~\ref{section:related} reviews related works and Section~\ref{section:conclusion} concludes this paper.

\section{Threat Model and Design Goals} \label{section:problem}

\subsection{Threat Model} \label{section:problem:threat}
We aim to develop a malicious traffic detection system as a plug-in module of middlebox. The middlebox forwards the replicated traffic to the detection system through port mirroring, which is similar to Cisco SPAN~\cite{SPAN}. Thus, the detection system does not interfere with benign traffic forwarding. We assume that the detection system does not have any prior knowledge on threats, which means that it should be able to deal with zero-day attacks~\cite{ACM-CS-16-survey, INFO20-ZeroWall, NDSS18-Kitsune}. Note that, we do not consider defenses against the attacks detected by \NM\ and can deploy existing malicious traffic defenses~\cite{NDSS20-Poseidon, USEC20-NetWarden, USEC21-Ripple} to throttle the detected traffic. 

The developed detection system should be able to determine whether traffic passing through the middlebox is benign or malicious by monitoring ongoing traffic. We emphasize that the malicious traffic detection is fully different from traffic classification~\cite{TIFS17-Meng, CCS17-Scalable-TA, NDSS20-DNS-FP, NDSS20-mobile-app-FP} that aims to classify whether traffic is generated by a certain network application or a certain user. We do not consider detecting passive attacks that do not cause obvious traffic variance, e.g., eavesdropping attacks and intercept attacks~\cite{NDSS12-ospf, ACSAC14-ospf}.

\subsection{Design Goals} \label{section:problem:goals}
In this paper, we aim to develop a realtime robust malicious traffic detection system, which achieves high detection accuracy and task-agnostic detection. Particularly, the system should achieve the following two goals, which are not well addressed in the literature. 

\noindent \textbf{Robust Accurate Detection.} The system should be able to detect various zero-day attacks. Especially, it can capture different evasion attacks, which try to evade detection by deliberately injecting noise packets, i.e., using various packets generated by benign applications, into the attack traffic. 

\noindent \textbf{Realtime Detection with High Throughput.} The system should be able to be deployed in high throughput networks, e.g., a 10 Gigabit Ethernet, while incurring low detection latency.

\section{Overview of \NM} \label{section:overview}
In this section, we present our malicious traffic detection system, \NM. \NM\ achieves high performance detection by encoding per-packet feature sequences as vectors to reduce the overhead of subsequent feature processing. Meanwhile, it extracts the sequential information of traffic via frequency domain to ensure detection accuracy. In particular, since the frequency domain features represent fine-grained sequential information of the packet sequences, which are not disturbed by the injected noise packets, \NM\ can achieve robust detection. Figure~\ref{diagram:high-level} shows the overview of \NM.  

\noindent \textbf{High Speed Packet Parser Module.} High speed packet parser module extracts per-packet features, e.g., the packet length and arriving time interval, at high speed to ensure the processing efficiency in both training and detection phases. This module provides the per-packet feature sequences to the feature extraction module for extracting the frequency domain features and the automatic parameter selection module for determining the encoding vector. Note that, this module dose not extract specific application related features and thus \NM\ achieves task agnostic detection.

\noindent \textbf{Frequency Features Extraction Module.} In both training and detection phases, this module extracts the frequency domain features from the per-packet feature sequences. This module periodically polls the required information from the high speed packet parser module with a fixed time interval. After acquiring the extracted per-packet features, it encodes the per-packet feature sequences as vectors and extracts the sequential information via frequency domain. These features with low redundancy are provided for the statistical clustering module. However, it is difficult to extract the frequency domain features of traffic in high throughput networks in realtime because of the various complicated, irregular, and dynamic flow patterns~\cite{S&P10-Outside, CCS04-Sommer}. We cannot apply deep learning models, e.g., recurrent neural networks, to extract features due to their long processing latency though they can extract more richer features for detection. We will present the details of this module in Section~\ref{section:detail:extract}.

\noindent \textbf{Automatic Parameter Selection Module.} This module calculates the encoding vector for the feature extraction module. We decide the encoding vector by solving a constrained optimization problem that reduces the mutual interference of different per-packet features. In the training phase, this module acquires the per-packet feature sequences and solves an equivalent Satisfiability Modulo Theories (SMT) problem to approximate the optimal solution of the original problem. By enabling automatic parameter selection, we significantly reduce the manual efforts for parameter selection. Therefore, we can fix and accurately set the encoding vector in the detection phase. We will describe the details of the module in Section~\ref{section:detail:select}.

\noindent \textbf{Statistical Clustering Module.} In this module, we utilize a lightweight statistical clustering algorithm to learn the patterns of the frequency domain features from the feature extraction module. In the training phase, this module calculates the clustering centers of the frequency domain features of benign traffic and the averaged training loss. In the detection phase, this module calculates the distances between the frequency domain features and the clustering centers. \NM\ detects traffic as malicious if the distances are significantly larger than the training loss. We will elaborate on the statistical clustering based detection in Section~\ref{section:detail:clustering}. \\

\section{Design Details} \label{section:detail}
In this section, we present the design details of \NM, i.e., the design of three main modules in \NM.  

\renewcommand{\arraystretch}{1.0}
\begin{figure*}[t]
    \subfigcapskip=-2mm
    \centering
	    \vspace{-2mm}
	    \subfigure[Benign TLS traffic and side-channel attack traffic]{ 
		    \includegraphics[width=0.33\textwidth]{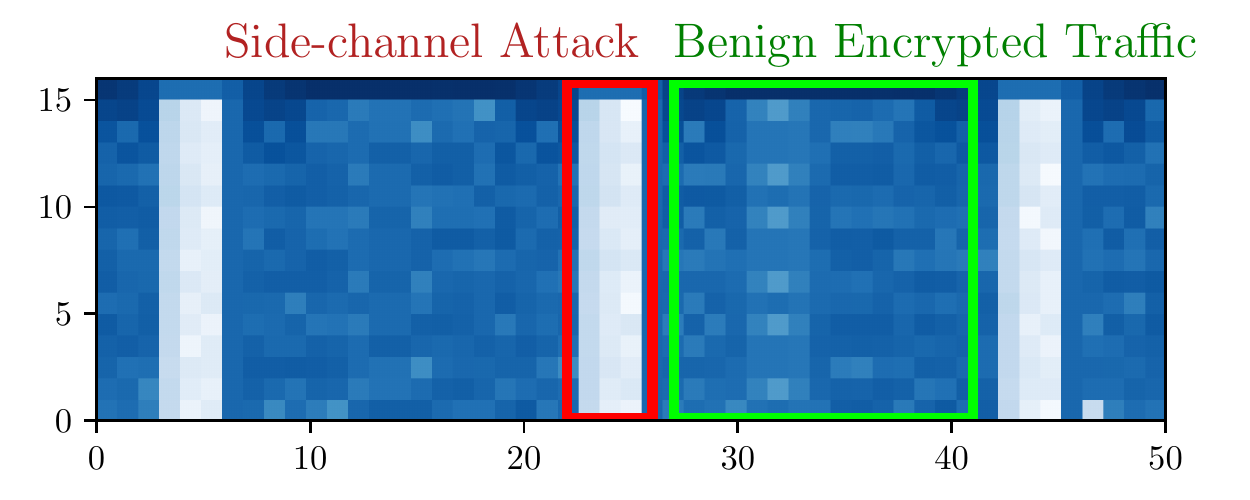}
	    }
	    \hspace{-4mm}
	    \subfigure[Benign UDP traffic and SSL DoS traffic]{
		    \includegraphics[width=0.33\textwidth]{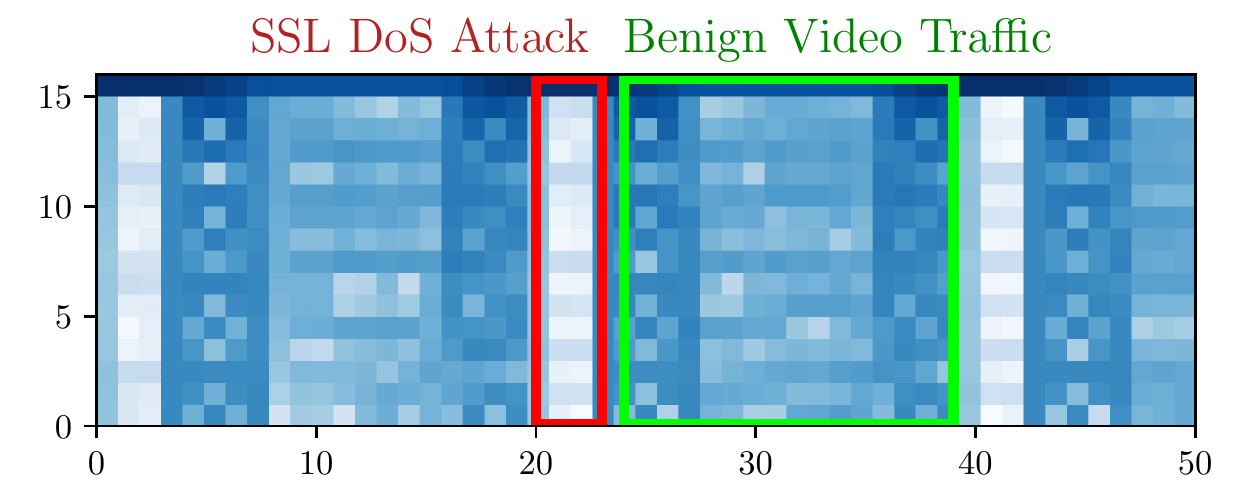}
	    }
	    \hspace{-4mm}
	    \subfigure[Outbound NAT traffic and LowRate TCP DoS traffic]{
		    \includegraphics[width=0.33\textwidth]{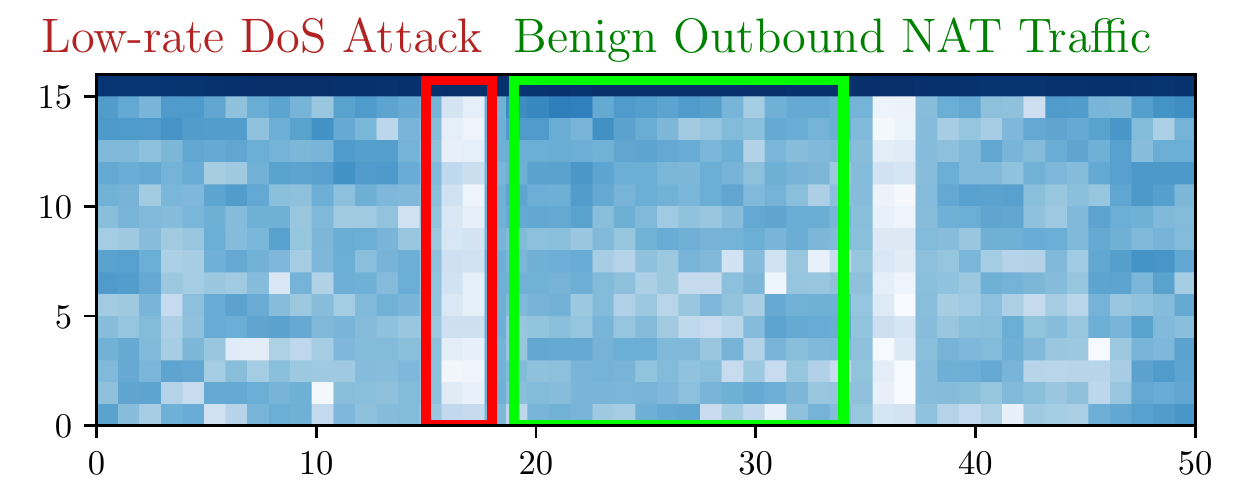}
		    \label{graph:visual:extrem2}
	    }
	\vspace{-4mm}
	\caption{We map the frequency domain features, which are extracted from the traffic with three types of typical attacks, to the RGB space, and observe that a small number of malicious packets incur significant changes in the frequency domain features.}
	\label{graph:visual}
	\vspace{-2mm}
\end{figure*}

\newcommand{\wsamp}{W_{win}}
\newcommand{\wseg}{W_\mathrm{seg}}

\subsection{Frequency Feature Extraction Module} \label{section:detail:extract}
In this module, we extract the frequency domain features from high speed traffic. We acquire the per-packet features of $N$ packets from the same flow by polling the high speed packet parser module. We use the mathematical representation similar to Bartos~\etal~\cite{USEC16-Variants} to denote the features. We use $s^{(i)}$ and $M$ to indicate the $i^{th}$ per-packet feature and the number of per-packet features, respectively. Matrix $\mathrm{S}$ denotes the per-packet features of all packets, where $\mathrm{s}_{ik}$ is defined as $i^{th}$ packet's $k^{th}$ property:
\begin{equation}
    \mathrm{S} = [s^{(1)}, \dots, s^{(i)}, \dots, s^{(M)}] =
    \begin{bmatrix}
        s_{11} & \cdots & s_{1M} \\
        \vdots & \ddots  & \vdots \\
        s_{N1} & \cdots & s_{NM}
    \end{bmatrix}.
\end{equation}

\noindent \textbf{Packet Feature Encoding.} We perform a linear transformation $w$ on $\mathrm{S}$ to encode the features of a packet to a real number $v_i$. $v$ denotes the vector representation of traffic:
\begin{equation}
    v = \mathrm{S}w = [v_1, \dots, v_i, \dots, v_N]^\mathrm{T}, \quad v_i = \sum_{k=1}^{M} s_{ik}w_k.
\end{equation}
The feature encoding reduces the scale of features, which significantly reduces the processing overhead of \NM. In Section~\ref{section:detail:select}, we will describe how \NM\ automatically selects parameters for the encoding vector $w$.

\noindent \textbf{Vector Framing.} Now we segment the vector representation with the step length of $\wseg$. The goal of segmentation is to reduce the complexity of the frequency domain features by constraining the long-term dependence between packets. If the frames are excessively long, the frequency domain features will become too complex to learn in the statistical learning module. $N_f$ denotes the number of the frames. We obtain the following equations:
\begin{equation}
    f_i = v[\hspace{-2pt}[(i - 1) \times \wseg : i \times \wseg]\hspace{-2pt}] \qquad (1 \le i \le N_f),
\end{equation}
\begin{equation}
    N_f = \left\lfloor\frac{N}{W_\mathrm{seg}} \right\rfloor.
\end{equation}
    
\noindent \textbf{Discrete Fourier Transformation.} In the next step, we perform the Discrete Fourier Transformation (DFT) on each frame $f_i$ to extract the sequential information via frequency domain and reduce the information loss incurred by the flow-level methods. We can acquire the frequency features of each frame as follows:\footnote{$j$ denotes an imaginary number.}
\begin{equation}
    F_i = \mathscr{F}(f_i) \quad (1 \le i \le N_f),
\end{equation}
\begin{equation}
    F_{ik}=\sum_{n=1}^{\wseg} f_{in} e^{-j\frac{2\pi(n-1)(k-1)}{\wseg}} \quad (1 \le k \le \wseg),
\end{equation}
where $F_{ik}$ is a frequency component of $i^{th}$ frame with the frequency of $2\pi(k-1) / \wseg$. Note that, all frequency features output by DFT are vectors with complex numbers, which cannot be used directly as the input for machine learning algorithms. 

\noindent \textbf{Calculating the Modulus of Complex Numbers.} We transform the complex numbers to real numbers by calculating the modulus for the frequency domain representation. For simplicity, we transform $F_{ik}$ to a coordinate plane representation:
\begin{equation}
    F_{ik} = a_{ik} + j  b_{ik},
\end{equation}
\begin{equation}
    \begin{cases}
    a_{ik} &= \sum\limits_{n=1}^{\wseg} f_{in} \cos\frac{2\pi(n-1)(k-1)}{\wseg} \\
    b_{ik} &= \sum\limits_{n=1}^{\wseg} -f_{in} \sin\frac{2\pi(n-1)(k-1)}{\wseg}.
    \end{cases}
\end{equation}
We calculate the modulus for $F_{ik}$ as $p_{ik}$ in \eqref{eq:power}. For the $i^{th}$ frame, we select the first half of the modulus as vector $P_i$. Because the transformation results of DFT are conjugate, the first half and the second half are symmetrical. Thus, we can obtain: 
\begin{equation}\label{eq:power}
    p_{ik} = a_{ik}^2 + b_{ik}^2 \quad (1 \le k \le \wseg),
\end{equation}
\begin{equation}
    P_{i} = [p_{i1}, \dots, p_{iK_f}]^\mathrm{T} \quad (K_f = \left\lfloor\frac{\wseg}{2}\right\rfloor + 1),
\end{equation}
\begin{equation}
    F_{ik} = F^{*}_{i(\wseg - k)} \quad \Rightarrow \quad p_{ik} = p_{i(\wseg - k)}.
\end{equation}

\noindent \textbf{Logarithmic Transformation.} To make the frequency domain features to be numerically stable~\cite{DeepLearning} and prevent float point overflow during the machine learning model training, we perform a logarithmic transformation on $P_i$, and use constant $C$ to adjust the range of the frequency domain features:
\begin{equation}
    R_i = \frac{\ln(P_i + 1)}{C} \quad (1 \le i \le N_f),
\end{equation}
\begin{equation}
    \mathrm{R}_{K_f \times N_f} = [R_1, \dots, R_i, \dots, R_{N_f}].
\end{equation}
As the output of the features extraction module, the $i^{th}$ column component of $\mathrm{R}$ is the frequency domain features of the $i^{th}$ frame. Matrix $\mathrm{R}$ is the input for the statistical clustering module.

Take an example, we collect three types of benign traffic (90\%) mixed with the malicious traffic (10\%) in Wide Area Network (WAN). We select 1500 continuous packets ($N=1500$) from each type of traffic and extract three per-packet features ($M=3$) including the packet length, the protocol type, and the arriving time interval. We fix the framing length $\wseg=30$. Therefore, $N_f=50$ and $K_f=16$. Then we perform a min-max normalization operation on the frequency domain features $\mathrm{R}$ and map the results to the RGB space. We visualize the frequency domain features that are similar to the Spectrogram in speech recognition~\cite{ICML16-DeepSpeech2}. As shown in Figure~\ref{graph:visual}, we observe that the area associated with the frequency domain features of the malicious traffic is significantly lighter than that of the benign traffic.

\subsection{Automatic Parameters Selection Module}\label{section:detail:select}
Now we determine the encoding vector $w$ for the feature extraction module that uses $w$ to encode the per-packet feature sequences and acquires the vector representation of the traffic. In general, we formulate the encoding vector selection problem as a constrained optimization problem, and transform the original problem into an equivalent SMT problem. We  approximate the optimal solution of the original problem through solving the SMT problem.

We assume that we can find a set of continuous functions to describe the changes of each kind of the per-packet feature $s^{(i)}$. Thus, we consider all obtained per-packet features are the samples of the continuous functions, which are denoted as $h_i(t)$ $(1 \le i \le M)$. We need to find a vector $w$ to amplify and superpose all these functions. Our key optimization objective is to minimize mutual interference and bound the overall range when superposing the functions. We can first bound the range of encoding vector $w$ and the range of the superposition function in the following:
\begin{equation}
    W_{min} \le w_i \le W_{max} \quad (1 \le i \le M),   \label{eq:constrain1}
\end{equation}
\begin{equation}
    \sum\limits_{i=1}^{M} w_i h_i(t) \le B,  \label{eq:constrain2}
\end{equation}
where $W_{min}$, $W_{max}$, $B$ are constants. We constrain the order preserving properties of the functions to ensure that different types of per-packet features do not interfere with each other when the feature extraction module performs packet encoding:
\begin{equation}
    w_i h_i(t) \le w_{i+1}h_{i+1}(t) \quad (1 \le i \le M -1) \label{eq:constrain3}.
\end{equation}

Second, we optimize $w$ to maximize the distances between the functions so that we can minimize the mutual interference of the per-packet features and bound the ranges of all the functions. Therefore, under the constrains of \eqref{eq:constrain1} \eqref{eq:constrain2} \eqref{eq:constrain3}, we obtain the optimization object:
\begin{multline}
    \qquad \hat{w} = \arg\max \int_{0}^{+\infty} w_M h_M(t) - w_1 h_1(t) \mathrm{d}t \quad - \\ 
    \sum\limits_{i=2}^{M-1} \int_{0}^{+\infty} \lvert 2w_i h_i(t) - w_{i+1}h_{i+1}(t) - w_{i-1}h_{i-1}(t) \lvert \mathrm{d}t. \label{eq:obj1}
\end{multline}

In practice, we cannot determine the convexity of the optimization object because the closed-form representations of $h_i(t)$ are not available. Thus, we reform the origin constrained optimization problem to a Satisfiability Modulo Theories (SMT) problem \eqref{eq:all-constrain} with optimization object \eqref{eq:obj2} to approximate the optimal solution of \eqref{eq:obj1}. For the $i^{th}$ per-packet feature, we perform a min-max normalization on $s_i$ and use $n_i$ to indicate the normalized vector. We list constrains \eqref{eq:all-constrain}. And we obtain the satisfied (SAT) solutions of the SMT problem and maximize the following objective:
\begin{multline}
    \qquad \widetilde{w} = \arg\max \sum\limits_{k=1}^{N} w_M n_{Mk}-w_1 n_{1k} \quad - \\
    \sum\limits_{i=2}^{M-1} 2w_i n_{ik} - w_{i-1} n_{(i-1)k} - w_{i+1} n_{(i+1)k}, \label{eq:obj2}
\end{multline}
subjects to:
\begin{equation}
    \begin{cases}
    w_i &\in \qquad [W_{min}, W_{max}] \\
    \sum\limits_{i=1}^{M} w_i n_{ik} &\le \qquad B \\
    w_i n_{ik} &\le \qquad w_{i+1} n_{(i+1)k} \\
    2w_i n_{ik} &\le \qquad w_{i-1} n_{(i-1)k} + w_{i+1} n_{(i+1)k}.
    \end{cases}
    \label{eq:all-constrain}
\end{equation}

Note that, we reform the absolute value operation in the optimization object \eqref{eq:obj1} into constrains \eqref{eq:all-constrain} because most SMT solvers do not support absolute value operations.

\subsection{Statistical Clustering Module} \label{section:detail:clustering}
Now we utilize the statistical clustering algorithm to learn the patterns of the frequency domain features obtained from the feature extraction module with the selected parameters. We train the statistical clustering algorithm with only benign traffic. In the training phase, this module calculates the clustering centers of the frequency domain features and the averaged training loss. In order to improve the robustness of \NM\ and reduce false positive caused by the extreme values, we segment the frequency domain feature matrix $\mathrm{R}$ with a sampling window of length $\wsamp$. We use $N_t$ to denote the number of samples and $l$ to denote the start points. We average the sampling window on the dimension of the feature sequence and use $r_i$ to indicate the input of the clustering algorithm. We can obtain: 
\begin{equation}
    l = i \wsamp \quad (0 \le i < N_t), \quad N_t =  \left\lfloor\frac{N_f}{\wsamp} \right\rfloor,
\end{equation}
\begin{equation}
    r_i = \mathsf{mean}(\mathrm{R}[\hspace{-2pt}[l : l + \wsamp]\hspace{-2pt}]).
\end{equation}

We perform the statistical clustering algorithm and acquire all clustering centers to represent the benign traffic patterns. We use $C_k$ to denote the $K_C$ clustering centers, where $(1 \le k \le K_C)$, and then calculate the averaged training loss.  For each $r_i$, we find the closest clustering center as $\hat{C_i}$ and we take averaged L2-norm as the training loss:
\begin{equation}
    \hat{C_i} =  \mathop{\arg\min}_{C_k}  \left\| C_k - r_i \right\|_2 \quad (1 \le i \le N_t),
\end{equation}
\begin{equation}
    \mathsf{train\_loss} = \frac{1}{N_t} \sum^{N_t}_{i=1} \left\| r_i - \hat{C_i} \right\|_2.
\end{equation}

In the detection phase, this module calculates the distances between the frequency domain features of traffic and the clustering centers. For each given frequency domain feature, we sample $N_t$ segments on $\mathrm{R}$ with length $\wsamp$, which is the same as the training phase. We can find the closest clustering center $\hat{C_i}$ as an estimate of $r_i$. We calculate the L2-norm as the estimation error:
\begin{equation}
    \mathsf{loss}_i = \mathsf{min}(\left\| r_i - C_k \right\|_2) \quad (1 \le k \le K_C).
\end{equation}

If the estimation error $\mathsf{loss}_i \ge (\phi \times \mathsf{train\_loss})$, we can conclude that the statistical clustering algorithm cannot understand the frequency domain features of the traffic, which means the traffic is malicious.

\section{Theoretical Analysis}\label{section:analysis}
In this section, we conduct a theoretical analysis to prove that \NM\ achieves lower information loss in feature extraction than the packet-level and the traditional flow-level methods, which ensures that \NM\ extracts traffic features accurately. Due to the page limitations, all proofs can be found in Appendix~\ref{section:appendix:1} -~\ref{section:appendix:5-6}. Moreover, we analyze the scale of the frequency domain features and the algorithmic complexity of \NM.

\subsection{Information Loss in \NM}\label{section:analysis:model}

\noindent \textbf{Traffic Feature Differential Entropy Model.}
First, we develop the traffic feature differential entropy model, a theoretical analysis framework that \textit{evaluates the efficiency of traffic features by analyzing the information loss incurred by feature extractions} from an information theory perspective~\cite{Entropy}. The framework aims to (i) model an observable packet-level feature as a stochastic process and observed features extracted from ongoing packets as the state random variables of the process; (ii) model feature extraction methods as algebraic transformations of the state random variables; (iii) evaluate the efficiency of the features by measuring the information loss during the transformations.

We model a particular type of packet-level feature (e.g., the packet length, and the time interval) as a discrete time stochastic process $\mathcal{S}$, which is used to model traffic feature extraction by different detection methods. We use a random variable vector $\vec{s}=[s_1, s_2, \dots, s_N]$ to denote a packet-level feature sequence extracted from $N$ continuous packets, i.e., $N$ random variables from $\mathcal{S}$. $\mathsf{f}$ indicates a feature extraction function that transforms the original features $\vec{s}$ for the input of machine learning algorithms. According to Table~\ref{table:packet-flow}, in the packet-level methods, $\mathsf{f}$ outputs the per-packet features sequence $\vec{s}$ directly. In the traditional flow-level methods, $\mathsf{f}$ calculates a statistic of $\vec{s}$. In \NM, $\mathsf{f}$ calculates the frequency domain features of $\vec{s}$. We assume that $\mathcal{S}$ is a discrete time Gaussian process, i.e., $\mathcal{S}\sim\mathrm{GP}(u(i), \Sigma(i, j))$. For simplicity, we mark $\Sigma(i, i)$ as $\sigma(i)$. We assume $\mathcal{S}$ is an independent process and then we can obtain the covariance function of $\mathcal{S}$, i.e., $\kappa(x_i, x_j)=\sigma(i)\delta{(i,j)}$. $p_i$ denotes the probability density function of $s_i$. We use differential entropy~\cite{Entropy} to measure the information in the features using the unit of \textit{nat}: 
\begin{equation} \label{eq:entropy}
\mathcal{H}(s_i) = - \int_{-\infty}^{+\infty} p_i(s) \ln p_i(s) \mathrm{d}s = \ln K\sigma(i), 
\end{equation}
where $K=\sqrt{2\pi e}$. We assume that the variance of each $s_i$ is large enough to ensure the significant change because a kind of stable packet-level feature is meaningless to be extracted and analyzed. Thus, we establish non-negative differential entropy assumption, i.e.,  $\sigma(i)\ge K^{-1}$ to ensure $\mathcal{H}(s_i) \ge 0$.

\noindent \textbf{Analysis of Traditional Flow-level Detection Methods.} We analyze the information loss in the feature extraction of the traditional flow-level methods. We consider three types of widely used statistical features in the traditional flow-level methods~\cite{TIFS19-Vesper, TIFS18-Qi, RAID19-SCADA, ACSAC12-Disclose, USEC08-Botminer}: (i) min-max features, the feature extraction function $\mathsf{f}$ outputs the maximum or minimum value of $\vec{s}$ to extract flow-level features of traffic and produces the output for machine learning algorithms. (ii) average features, $\mathsf{f}$ calculates the average number of $\vec{s}$ to obtain the flow-level features. (iii) variance features, $\mathsf{f}$ calculates the variance of $\vec{s}$ for machine learning algorithms. We analyze the information loss when performing the statistical feature extraction function $\mathsf{f}$. Based on the probability distribution of the state random variables and Equation~\eqref{eq:entropy}, we obtain the information loss of flow-level statistical features in the traditional flow-level detection over the packet-level detection and have the following properties of the features above.

\textbf{Theorem 1.} \textit{(The Lower Bound for Expected Information Loss of the Min-Max Features).}
For the min-max statistical features, the lower bound of expected information loss is:
\begin{equation}
    \mathrm{E}[\Delta \mathcal{H}_{\mathrm{flow-minmax}}] \ge (N -1 )\ln K \mathrm{E}[\sigma].
\end{equation}

\textbf{Theorem 2.} \textit{(The Lower Bound for Expected Information Loss of the Average Features).} The lower bound for the expectation of information loss in the average features is: 
\begin{equation}
    \mathrm{E}[\Delta \mathcal{H}_{\mathrm{flow-avg}}] \ge \ln \sqrt{N} K^{N-1} \mathrm{E}[\sigma]^{N-1}.
\end{equation}

We can obtain that the equality of Theorem 1 and Theorem 2 holds iff the stochastic process $\mathcal{S}$ is strictly stationary.  

\textbf{Theorem 3.} \textit{(The Lower Bound and Upper Bound for the Information Loss of the Average Features).}
For the average features, the upper and lower bounds of the information loss in the metric of differential entropy is:
\begin{equation}
    \ln N \le \Delta \mathcal{H}_{\mathrm{flow-avg}} \le \ln \sqrt{N} K^{N - 1} Q(\sigma)^{N - 1}, 
\end{equation}
where $Q(\sigma)$ is the square mean of the variances of the per-packet features sequence $\vec{s}$. 

\textbf{Theorem 4.} \textit{(The Information Loss of the Variance Features).}
When the Gaussian process $\mathcal{S}$ is strictly stationary with zero mean, i.e., $u(i)=0$ and $\sigma(i)=\sigma$, for the variance features, an estimate of the information loss is:
\begin{equation}
    \Delta \mathcal{H}_{\mathrm{flow-var}} = N \ln K\sigma - \ln \frac{\sqrt{4\pi N^3}}{\sigma^2}. 
\end{equation}

According to the theorems above, we can conclude that the information loss in the traditional flow-level detection methods increases approximately linearly with the length of per-packet feature sequences. Thus, comparing with the packet-level methods, the traditional flow-level methods cannot effectively extract the features of traffic. Although the traditional flow-level methods can adopt multiple statistical features~\cite{USEC16-Variants, KDD13-Active}, the number of packets in the feature extraction ($N$) is significantly larger than the number of features. In Section~\ref{section:evaluation:accuracy}, we will use experiments to show that the traditional flow-level methods achieve low detection accuracy.

\noindent \textbf{Analysis of \NM.} Different from the traditional flow-level methods, \NM\ encodes per-packet features as vectors and performs DFT on the vectors to extract the frequency domain features of the traffic. We prove the low information loss property of \NM\ by comparing with the packet-level methods (see Theorem 5) and the traditional flow-level methods (see Theorem 6) by leveraging the bounds of the information loss in Theorem 1 - 4.

\textbf{Theorem 5.} \textit{(An Estimation of the Information Loss of \NM\ over the Packet-level Methods).} When the Gaussian process $\mathcal{S}$ is strictly stationary with zero mean, i.e., $u(i)=0$ and $\sigma(i)=\sigma$, we can acquire an estimate of the information loss in \NM\ when ignoring the logarithmic transformation using:
\begin{equation}
    \Delta \mathcal{H}_{\mathrm{\NMM}} = N\ln \frac{\sigma}{w_i^2} \sqrt{\frac{\pi}{2e}} - N \ln N,
\end{equation}
where $w_i$ is the $i^{th}$ element of the encoding vector $w$. 

\textbf{Theorem 6.} \textit{(An Estimation of the Information Loss Reduction of \NM\ over the Traditional Flow-level Methods).} With the same assumption in Theorem 5, compared with the traditional flow-level methods that extract the average features, \NM\ reduces the information loss with an estimation:
\begin{align}
    \Delta \mathcal{H}_{\mathrm{\NMM-avg}} &= \Delta \mathcal{H}_{\mathrm{flow-avg}} - \Delta \mathcal{H}_{\mathrm{\NMM}} \\
                                                &= N\ln 2ew_i^2N + \ln \frac{\sqrt{N}}{K\sigma}.
\end{align}
Similarly, \NM\ reduces the information loss in the flow-level methods that use min-max features and variance features. We present the estimations of reduced information loss in the metric of differential entropy as follows:
\begin{equation}
    \Delta \mathcal{H}_{\mathrm{\NMM-minmax}} = N\ln 2ew_i^2N - \ln K\sigma,
\end{equation}
\begin{equation}
    \Delta \mathcal{H}_{\mathrm{\NMM-var}} = N\ln 2ew_i^2N - \ln \frac{\sqrt{4\pi N^3}}{\sigma^2} .
\end{equation}

According to Theorem 5, by using the packet-level methods as a benchmark, we conclude that \NM\ almost has no information loss when the number of packets involved in feature extraction is large. Thus, the feature efficiency of \NM\ is not worse than the packet-level methods. Moreover, the packet-level methods have a large feature scale that results in high overhead for machine learning (proof in Section~\ref{section:analysis:scale}). 

Based on Theorem 6, we conclude that the reduction of the information loss in the traditional flow-level methods increases more than linearly. Thus, by reducing the information loss in the traditional flow-level methods, \NM\ can extract features from ongoing traffic more effectively than the traditional flow-level methods. In Section~\ref{section:evaluation:accuracy}, we will measure the detection accuracy improvement of \NM\ by using experiments.

\subsection{Analysis of Scalability and Overhead}\label{section:analysis:scale}
\noindent \textbf{Feature Scale Reduction of \NM.} Original per-packet features are compressed in \NM. \NM\ reduces the input data scale and the processing overhead in machine learning algorithms. The compressed frequency domain features allow us to apply the machine learning algorithm in high throughput networks in practice. Compared with the packet-level methods, \NM\ achieves high compression ratio $C_r$ with a theoretical lower bound:

\begin{equation}
    C_r = \frac{\mathsf{size}(\mathrm{R})}{\mathsf{size}(\mathrm{S})} = \frac{K_f N_f}{MN} \ge \frac{(\frac{N}{\wseg})(\frac{\wseg}{2} + 1)}{MN} \ge \frac{1}{2M}.
\end{equation}

By reducing the feature scale, \NM\ significantly reduces the processing overhead in the packet-level methods and achieves high throughput. In  Section~\ref{section:evaluation:runtime}, we will show the experimental results of \NM\ to validate the analysis results. 

\noindent \textbf{Overhead of Feature Extraction in \NM.} \NM\ incurs a low computational overhead of extracting the frequency domain features from traffic. Particularly, \NM\ does not have an operation with high time or space complexity that is higher than quadratic terms. The time complexity and space complexity of \NM\ are shown in Table~\ref{table:complex}.

According to Table~\ref{table:complex}, the computational complexity of \NM\ is proportional to the number of packets $N$. Most of the consumption is incurred by matrix multiplications in the packet encoding. Compared with the encoding, performing DFT on frames has relatively less computation overhead and consumes less memory space because of the high speed DFT operation, i.e., Fast Fourier Transformation (FFT). In Section~\ref{section:evaluation:runtime}, we will validate the complexity of \NM\ by using the experimental results.

\renewcommand{\arraystretch}{1.0}
\begin{table}[t]
    \small
    \caption{Complexity of the Feature Extraction Module}
    \vspace{-4mm}
    \begin{center}
        \begin{tabular}{c|c|c}
        \toprule
        \textbf{Steps} & \textbf{Time Complexity} & \textbf{Space Complexity} \\
        \midrule
        Packet Encoding         & $O(MN)$               & $O(MN)$   \\
        Vector Framing          & $O(1)$                & $O(1)$        \\
        DFT Transformation      & $O(N\log\wseg)$       & $O(\wseg)$    \\
        Calculating Modulus     & $O(N/2)$              & $O(N)$        \\
        Log Transformation      & $O(N/2)$              & $O(1)$        \\
        \midrule
        Total                   & $O(MN + N\log\wseg)$  & $O(MN+\wseg)$ \\
        \bottomrule
        \end{tabular}
    \label{table:complex}
    \end{center}
    \vspace{-1mm}
\end{table}

\section{Experimental Evaluation} \label{section:evaluation}
In this section, we prototype \NM\ and evaluate its performance by using 42 real-world attacks. In particular, the experiments will answer the three questions:
\begin{enumerate}
    \item If \NM\ achieves higher detection accuracy than the state-of-the-art method? (Section~\ref{section:evaluation:accuracy})
    \item If \NM\ is robust to detect attacks even if an attackers try to evade the detection of \NM\ by leveraging the benign traffic? (Section~\ref{section:evaluation:robustness})
    \item If \NM\ achieves high detection throughput and low detection latency? (Section~\ref{section:evaluation:runtime})
\end{enumerate}

\subsection{Implementation}\label{section:evaluation:implementation}
We prototype \NM\ using C/C++ (GCC version 5.4.0) and Python (version 3.8.0) with more than 3,500 lines of code (LOC). The source code of \NM\ can be found in~\cite{Code-release}.

\noindent {\bf High Speed Packet Parser Module.} We leverage Intel Data Plane Development Kit (DPDK) version 18.11.10 LTS ~\cite{DPDK} to implement the data plane functions and ensure high performance packet parsing in high throughput networks. We bind the threads of \NM\ on physical cores using DPDK APIs to reduce the cost of context switching in CPUs. As discussed in Section~\ref{section:detail:extract}, we parse the three per-packet features, i.e., lengths, timestamps, and protocol types.

\noindent {\bf Frequency Domain Feature Extraction Module.} We leverage PyTorch~\cite{PyTorch} (version 1.6.0) to implement matrix transforms (e.g., encoding and Discrete Fourier Transformation) of origin per-packet features and auto-encoders in baseline methods.

\noindent {\bf Statistical Clustering Module.} We leverage K-Means as the clustering algorithm with the mlpack implementation (version 3.4.0)~\cite{mlpack} to cluster the frequency domain features.

\noindent {\bf Automatic Parameter Selection.} We use Z3 SMT solver (version 4.5.1)~\cite{Z3} to solve the SMT problem in Section~\ref{section:detail:select}, i.e., determining the encoding vector in \NM.

Moreover, we implement a traffic generating tool using Intel DPDK to replay malicious traffic and benign traffic simultaneously. The hyper-parameters used in \NM\ are shown in Table~\ref{table:configure}.

\begin{table}[t]
\small
\centering
    \caption{Recommended Hyper-parameter Configurations}
    \vspace{-4mm}
    \begin{center}
        \begin{tabular}{@{}c|c|c@{}}
        \toprule
        \textbf{Hyper-Parameters}   & \textbf{Description}   & \textbf{Value} \\
        \midrule
        $\wseg$                     & Framing length                & 50                \\
        $\wsamp$                    & Sampling window length        & 100               \\
        $C$                         & Adjusting frequency domain features  & 10         \\
        $K_C$                       & Number of clustering centers  & 10                \\
        $[W_{min}, W_{max}]$        & Range of the encoding vector  & [10, $10^3$]      \\
        $B$                         & Upper bound of the encoded features & $10^5$      \\
        \bottomrule
        \end{tabular}
    \label{table:configure}
    \end{center}
    \vspace{-2mm}
\end{table}

\subsection{Experiment Setup}\label{section:evaluation:setup}

\noindent \textbf{Baselines.} To measure the improvements achieved by \NM, we establish three baselines:
\begin{itemize}

\item \textit{Packet-level Detection.} We use the state-of-the-art machine learning based detection method, Kitsune~\cite{NDSS18-Kitsune}. It extracts per-packet features via flow state variables and feeds the features to auto-encoders. We use the open source Kitsune implementation~\cite{Kitsune-py} and run the system with the same hardware as \NM.

\item \textit{Flow-level Statistics Clustering (FSC).}  As far as we know, there is no flow-level malicious traffic detection method that achieves task agnostic detection. Thus, we establish 17 flow-level statistics according to the existing studies~\cite{USEC16-Variants, RAID19-SCADA, ACSAC12-Disclose, TIFS19-Vesper, TIFS18-Qi, S&P20-Wang} including the maximum, minimum, variance, mean, range of the per-packet features in \NM, flow durations, and flow byte counts. We perform a normalization for the flow-level statistics. For a fair comparison, we use the same clustering algorithm to \NM.

\item \textit{Flow-level Frequency Domain Features with Auto-Encoder (FAE).} We use the same frequency domain features as \NM\ and an auto-encoder model with 128 hidden states and Sigmoid activation function, which is similar to the auto-encoder model used in Kitsune. For the training of the auto-encoder, we use the Adam optimizer and set the batch size as 128, the training epoch as 200, the learning rate as 0.01.
\end{itemize}

\renewcommand{\arraystretch}{1.0}
\begin{table*}[t]
    \small
    \caption{Attack Dataset Configurations}
    \vspace{-4mm}
    \begin{center}
        \begin{tabular}{c|c|c|l|cc|c}
        \toprule
        \textbf{\tabincell{c}{Group}} & \textbf{Label} & \textbf{Attack Description} & \textbf{\tabincell{c}{Benign\\Traffic$^{\rm 1}$}} & \textbf{\tabincell{c}{Benign\\Flow Rate}} & \textbf{\tabincell{c}{Malicious\\Flow Rate}} & \textbf{\tabincell{c}{Ratio of\\Malicious$^{\rm 2}$}} \\
        \midrule
        \multirow{6}{*}{\tabincell{c}{Traditional\\Attacks}} & SYN DoS & TCP SYN flooding Deny-of-Service attack. &  2020.6.10 & 5.276 Gbps & 23.04 Mbps & 0.0858 \\
        & Fuzz Scan & Scanning for vulnerabilities in protocols. & 2020.6.10 & 5.276 Gbps & 27.92 Mbps & 0.0089 \\
        & OS Scan & Scanning for active hosts with vulnerable operating systems. & 2019.1.2 & 4.827 Gbps & 0.960 Mbps & 0.0045 \\
        & SSL DoS & SSL renegotiation messages flooding Deny-of-Service attack. & 2020.1.1 & 7.666 Gbps & 21.60 Mbps & 0.0128 \\ 
        & SSDP DoS & SSDP flooding Deny-of-Service attack. &  2020.1.1 & 7.666 Gbps & 27.20 Mbps & 0.0321 \\
        & UDP DoS & High-rate UDP traffic blocks bottleneck links. & 2019.1.2 & 4.827 Gbps & 2.422 Gbps & 0.4712 \\ 
        \midrule
        \multirow{3}{*}{\tabincell{c}{Multi-stage\\TCP Attacks}} & IPID SC & Side-channel attack via IPID assignments, disclosed in 2020~\cite{CCS20-MySC}. & 2020.6.10 & 5.276 Gbps & 0.138 Mbps & 0.0007 \\
        & ACK SC & ACK rate limit side-channel attack, disclosed in 2016~\cite{USEC16-ACKSC}. & 2019.1.2 & 4.827 Gbps & 1.728 Mbps & 0.0091 \\ 
        & TLS Oracle & TLS padding oracle attack~\cite{EUROCRYPT02-TLS}. & 2020.1.1 & 7.666 Gbps & 1.626 Mbps & 0.0031 \\
        \midrule
        \multirow{5}{*}{\tabincell{c}{Stealthy\\TCP Attacks}} & LRDoS 0.2 & UDP burst triggers TCP retransmissions (burst interval 0.2s). & 2019.1.2 & 4.827 Gbps & 0.115 Gbps & 0.0228 \\
        & LRDoS 0.5 & UDP burst triggers TCP retransmissions (burst interval 0.5s). & 2019.1.2 & 4.827 Gbps & 0.046 Gbps & 0.0112 \\ 
        & LRDoS 1.0 & UDP burst triggers TCP retransmissions (burst interval 1.0s). & 2019.1.2 & 4.827 Gbps & 0.023 Gbps & 0.0055 \\
        & IPID Scan & Prerequisite scanning of the IPID side-channel attack~\cite{CCS20-MySC}. & 2020.6.10 & 5.276 Gbps & 0.214 Mbps & 0.0010 \\
        & TLS Scan & TLS vulnerabilities scanning~\cite{USEC19-SCAN}. & 2020.6.10 & 5.276 Gbps & 0.046 Gbps & 0.0071 \\
        \bottomrule
        \multicolumn{4}{l}{\footnotesize $^{\rm 1}$ The Benign Traffic column shows the identifier (date) of WIDE MAWI traffic datasets~\cite{WIDE}. \vspace{-0.8mm}} \\
        \multicolumn{4}{l}{\footnotesize $^{\rm 2}$ The Ratio of Malicious column shows the packet number ratio of benign and malicious traffic.}
        \end{tabular}
    \end{center}
    \label{table:dataset}
    \vspace{-4mm}
\end{table*}

\noindent \textbf{Testbed.} We conduct the \NM, FSC, and FAE experiments on a testbed built on a DELL server with two Intel Xeon E5645 CPUs (2 $\times$ 12 cores), Ubuntu 16.04 (Linux 4.15.0 LTS), 24GB memory, one Intel 10 Gbps NIC with two ports that supports DPDK, and Intel 850nm SFP+ laser ports for optical fiber connections. We configure 8GB huge page memory for DPDK (4GB/NUMA Node). We bind 8 physical cores for 8 NIC RX queues to extract per-packet features and the other 8 cores for \NM\ analysis threads, which extract the frequency domain features of traffic and perform statistical clustering. In summary, we use 17 of 24 cores to enable \NM. Note that, since Kitsune cannot handle high-rate traffic, we evaluate it with offline experiments on the same testbed. 

We deploy DPDK traffic generators on the other two servers with similar configurations. The reason why we use two traffic generators is that the throughput of \NM\ exceeds the physical limit of 10 Gbps NIC, i.e., 13.22 Gbps. We connect two flow generators with optical fibers to generate high speed traffic. 

\noindent \textbf{Datasets.} The datasets used in our experiments are shown in Table~\ref{table:dataset}. We use three recent datasets from the WIDE MAWI Gigabit backbone network~\cite{WIDE}. In the training phase, we use 20\% benign traffic to train the machine learning algorithms. We use the first 20\% packets in MAWI 2020.06.10 dataset to calculate the encoding vector via solving the SMT problem (see Section~\ref{section:detail:select}). Meanwhile, we replay four groups of malicious traffic combined with the benign traffic on the testbed: 

\begin{itemize}
\item \textit{Traditional DoS and Scanning Attacks.} We select five active attacks from the Kitsune~\footnote{We exclude passive attack datasets without malicious flow but only victim flow. Note that, in our threat model we do not consider attacks without malicious packets.}~\cite{NDSS18-Kitsune} and a UDP DoS attack trace~\cite{UDP-dataset} to measure the accuracy of detecting high-rate malicious flow. To further evaluate \NM, we collect new malicious traffic datasets on WAN including Multi-Stage TCP Attacks, Stealthy TCP Attacks, and Evasion Attacks.

\item \textit{Multi-Stage TCP Attacks.} TCP side-channel attacks exploit the protocol implementations and hijack TCP connections by generating forged probing packets. Normally, TCP side-channel attacks have several stages, e.g., active connection finding, sequence number guessing, and acknowledgement number guessing. We implement two recent TCP side-channel attacks~\cite{CCS20-MySC, USEC16-ACKSC}, which have different numbers of attack stages. Moreover, we collect another multi-stage attack, i.e., TLS padding oracle attack~\cite{EUROCRYPT02-TLS}.

\item \textit{Stealthy TCP Attacks.} The low-rate TCP DoS attacks generate low-rate burst traffic to trick TCP congestion control algorithms and slow down their sending rates~\cite{TON06-LRTCPDOS, SIGCOMM03-LRTCPDOS, NDSS10-LRTCPDOS}. Low-rate TCP DoS attacks are more stealthy than flooding based DoS attacks. We construct the low-rate TCP DoS attacks with different sending rates. Moreover, we replay other low-rate attacks, e.g., stealthy vulnerabilities scanning~\cite{USEC19-SCAN}. 

\item \textit{Evasion Attacks.}  We use evasion attack datasets to evaluate the robustness of \NM. Attackers can inject noise packets (i.e., benign packets of network applications) into malicious traffic to evade detection~\cite{evade2}. For example, an attacker can generate benign TLS traffic so that the attacker sends malicious SSL renegotiation messages and the benign TLS packets simultaneously. Basing on the typical attacks above, we adjust the ratio of malicious packets and benign packets, i.e., the ratio of 1:1, 1:2, 1:4, and 1:8, and the types of benign traffic to generate 28 datasets. For comparison, we replay the evasion attack datasets with the same background traffic in Table~\ref{table:dataset}.
\end{itemize}

\newcommand{\ap}{$\approx$}
\newcommand{\rt}[1]{\color[rgb]{0.15, 0.616, 0.15}#1}
\newcommand{\rb}[1]{\color[rgb]{0.753,0,0}#1}

\renewcommand{\arraystretch}{1.0}
\begin{table*}[!t]
    \small
    \centering
    \setlength\tabcolsep{4.4pt}
    \caption{Detection Accuracy of \NM\ and Baselines on 14 Attacks}
    \vspace{-4mm}
    \begin{center}
    \begin{tabular}{@{}c|cccc|cccc|cccc|cccc@{}}
    \toprule
        Methods &  \multicolumn{4}{|c|}{Kitsune} & \multicolumn{4}{|c|}{FSC} & \multicolumn{4}{|c|}{FAE} & \multicolumn{4}{|c}{\NMM} \\
        \midrule
        Metrics & TPR & FPR & AUC & EER & TPR & FPR & AUC & EER & TPR & FPR & AUC & EER & TPR & FPR & AUC & EER \\
        \midrule
        SYN DoS & \rb 0.9801 & \rb 0.0910 & \rb 0.9562 & \rb 0.0919 & \rt 0.9999 & 0.0396 & 0.9603 & 0.0396 & 0.9813 & \rt 0.0033 & 0.9840 & \rt 0.0186 & 0.9924 & 0.0329 & \rt 0.9870 & 0.0512  \\
        Fuzz Scan & 0.9982 & \rt 0.0015 & \rt 0.9978 & 0.0336 & 0.0000 & \rb 0.4007 & \rb 0.6028 & \rb 0.3964 & \rb 0.0000 & 0.4111 & 0.6134 & 0.3954 & \rt 0.9999 & 0.0046 & 0.9962 & \rt 0.0047 \\
        OS Scan & 0.9997 & 0.0786 & 0.9615 & 0.0800 & \rb 0.0000 & \rb 0.1114 & \rb 0.8885 & \rb 0.1114 & 0.9999 & \rt 0.0069 & 0.9907 & \rt 0.0075 & \rt 0.9999 & 0.0106 & \rt 0.9951 & 0.0111 \\
        SSL DoS & 0.9417 & \rt 0.0035 & \rt 0.9781 & 0.0574 & \rt 0.9992 & 0.0519 & 0.9732 & \rt 0.0519 & \rb 0.0000 & \rb 0.1271 & \rb 0.8774 & \rb 0.1271 & 0.9699 & 0.0796 & 0.9391 & 0.0798 \\
        SSDP DoS & 0.9901 & 0.0132 & 0.9955 & 0.0168 & \rt 0.9999 & 0.0014 & 0.9986 & \rt 0.0014 & \rb 0.0003 & \rb 0.1233 & \rb 0.8770 & \rb 0.1233 & 0.9969 & \rt 0.0117 & \rt 0.9902 & 0.0172 \\
        UDP DoS & \rb 0.4485 & \rb 0.1811 & \rb 0.8993 & \rb 0.1433 & 0.9999 & 0.0173 & 0.9826 & 0.0173 & 0.9999 & \rt 0.0068 & \rt 0.9942 & \rt 0.0071 & \rt 0.9999 & 0.0083 & 0.9922 & 0.0093 \\
        \midrule
        IPID SC & \rb / & \rb / & \rb / & \rb / & 0.0000 & 0.2716 & 0.7702 & 0.2716 & \rt 0.8913 & \rt 0.1001 & \rt 0.9739 & \rt 0.1001 & 0.6900 & 0.2324 & 0.9322 & 0.2014 \\
        ACK SC & \rb / & \rb / & \rb / & \rb / & 0.0000 & 0.3090 & 0.6909 & 0.3090 & \rb - & \rb - & \rb - & \rb - & \rt 0.9999 & \rt 0.0001 & \rt 0.9999 & \rt 0.0001 \\
        TLS Oracle & 0.9973 & 0.0335 & 0.9722 & 0.0392 & \rb - & \rb - & \rb - & \rb - & \rb - & \rb - & \rb - & \rb - & \rt 0.9999 & \rt 0.0121 & \rt 0.9885 & \rt 0.0124 \\
        \midrule
        LRDoS 0.2 & \rb 0.6397 & \rb 0.1270 & \rb 0.9202 & \rb 0.1239 & 0.9999 & 0.0254 & 0.9740 & 0.0254 & 0.9999 & 0.0254 & \rt 0.9925 & \rt 0.0088 & \rt 0.9999 & \rt 0.0109 & 0.9915 & 0.0123 \\
        LRDoS 0.5 & \rb 0.0208 & \rb 0.1882 & \rb 0.8480 & \rb 0.1835 & \rt 0.9999 & 0.0551 & 0.9448 & 0.0551 & 0.9999 & \rt 0.0078 & \rt 0.9925 & \rt 0.0081 & 0.9999 & 0.0101 & 0.9916 & 0.0114 \\
        LRDoS 1.0 & \rb 0.0015 & \rb 0.1774 & \rb 0.8373 & \rb 0.1758 & 0.9999 & 0.0940 & 0.9059 & 0.0940 & \rt 0.9999 & \rt 0.0074 & \rt 0.9935 & \rt 0.0074 & 0.9999 & 0.0115 & 0.9910 & 0.0122 \\
        IPID Scan & \rb - & \rb - & \rb - & \rb - & \rt 0.9999 & 0.0801 & 0.9255 & 0.0801 & 0.9999 & \rt 0.0155 & \rt 0.9934 & \rt 0.0179 & 0.7964 & 0.1601 & 0.9579 & 0.1259 \\
        TLS Scan & \rb - & \rb - & \rb - & \rb - & \rb - & \rb - & \rb - & \rb - & 0.0000 & 0.4014 & 0.6033 & 0.3973 & \rt 0.9999 & \rt 0.0091 & \rt 0.9905 & \rt 0.0095 \\
        \bottomrule
        \multicolumn{17}{l}{\footnotesize $^{\rm 1}$ We highlight the best in {\rt{$\bullet$}} and the worst in {\rb{$\bullet$}} and we mark - when AUC $<$ 0.5 (meaningless, no better than random guess). \vspace{-0.8mm}} \\
        \multicolumn{17}{l}{\footnotesize $^{\rm 2}$ We mark / when Kitsune cannot finish the detection in 2 hours due to a large number of maintained flow state variables (process speed $< 4\times 10^3$ PPS).}
    \end{tabular}
    \end{center}
    \label{table:comparsion}
    \vspace{-3mm}
\end{table*}

\noindent \textbf{Metrics.}
We use the following metrics to evaluate the detection accuracy: (i) true-positive rates (TPR), (ii) false-positive rates (FPR), (iii) the area under ROC curve (AUC), (vi) equal error rates (EER). Moreover, we measure the throughput and processing latency to demonstrate that \NM\ achieves realtime detection. 

\subsection{Detection Accuracy}\label{section:evaluation:accuracy}
\begin{figure}[t]
    \subfigcapskip=-2mm
    \vspace{-2mm}
    \begin{center}
	\subfigure[SSL DoS]{
		\includegraphics[width=0.23\textwidth]{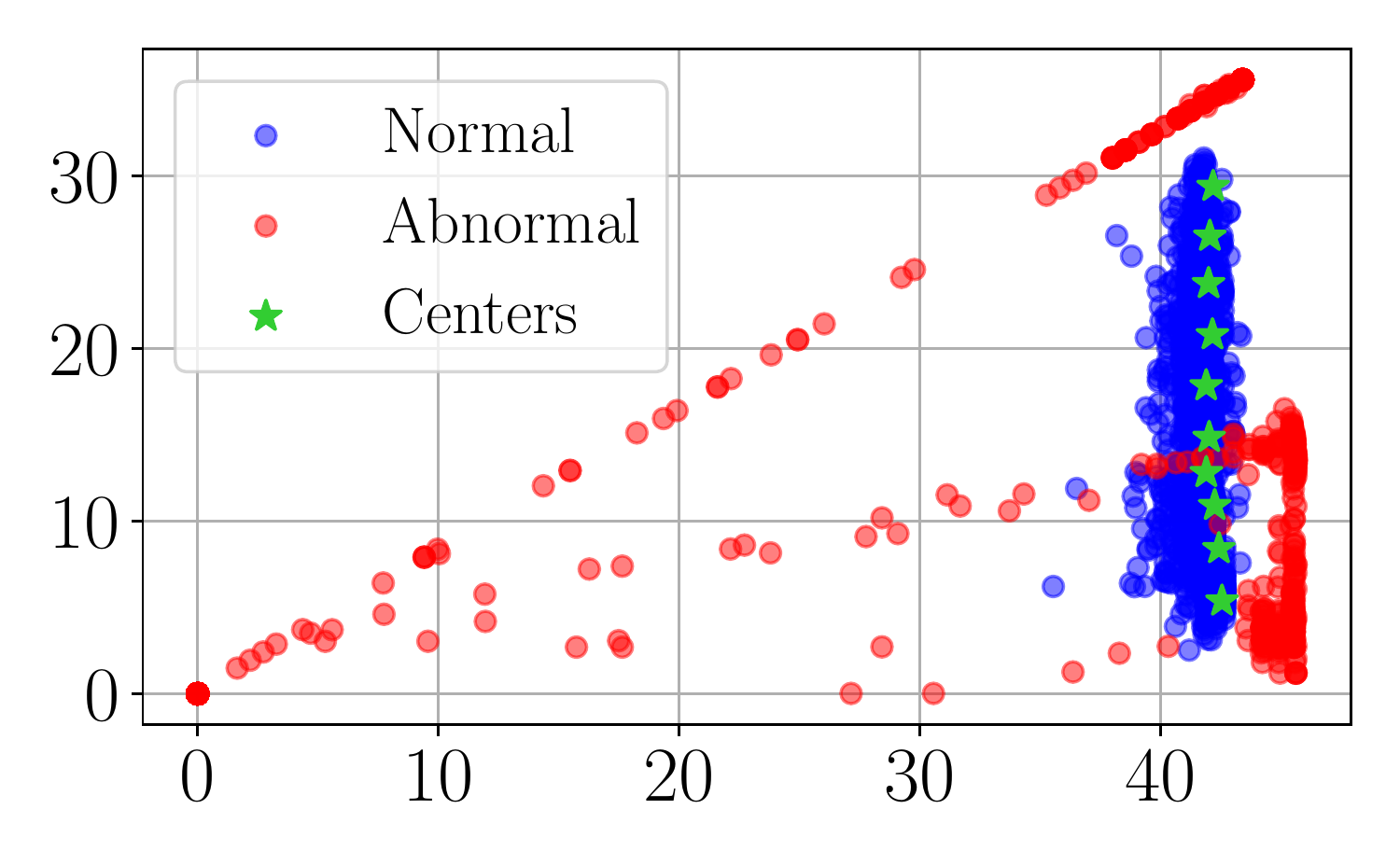}
	} 
	\hspace{-2mm}
	\subfigure [TLS Padding Oracle]{ 
		\includegraphics[width=0.23\textwidth]{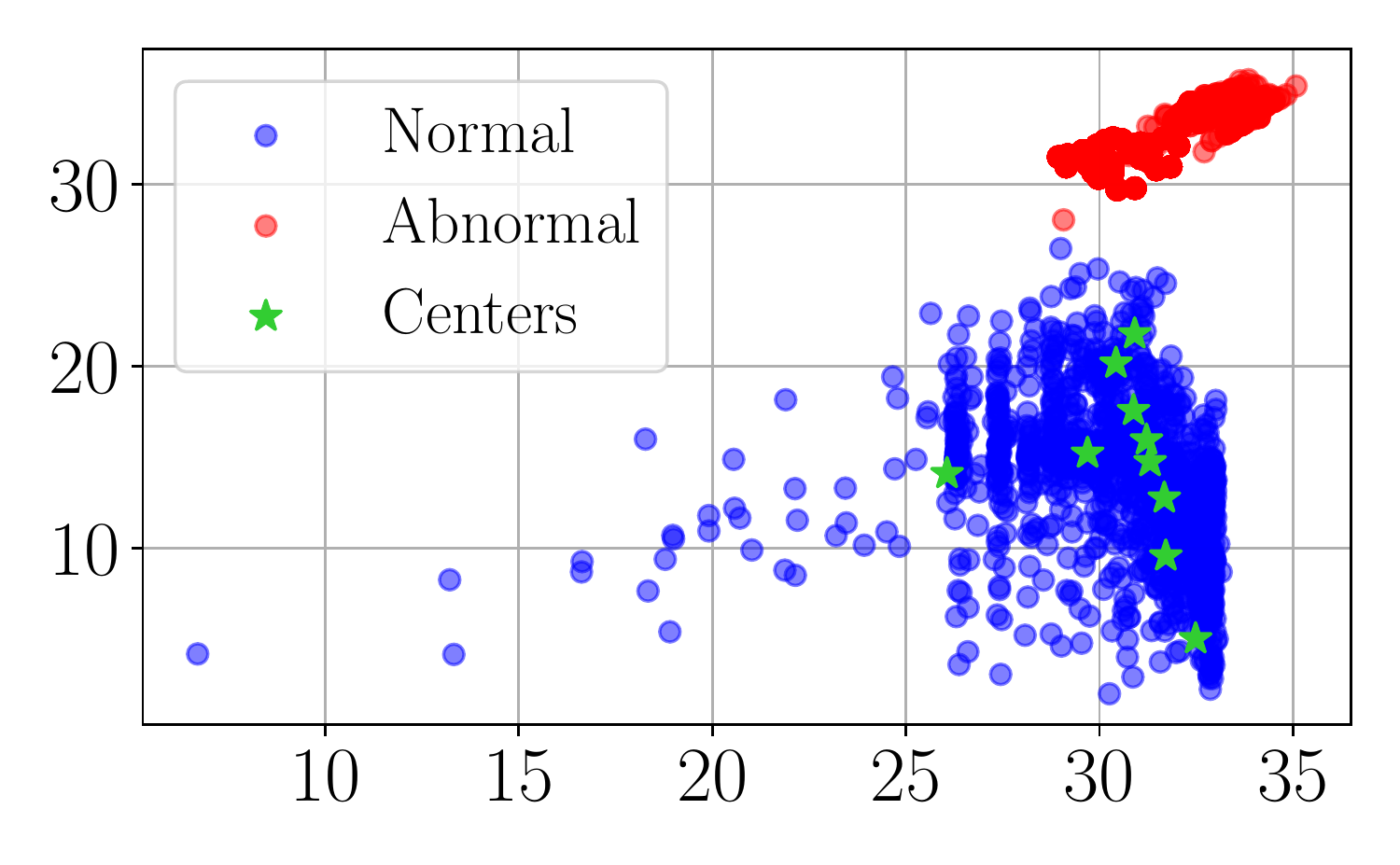}
	}
	\vspace{-4mm}
	\caption{Frequency domain features clustering results of \NM.} 
	\label{graph:center}
	\end{center}
	\vspace{-3mm}
\end{figure}

\begin{figure}[t]
    \subfigcapskip=-2mm
    \vspace{-2mm}
    \begin{center}
	\subfigure[SYN Flooding DoS (23.04 Mbps)]{
		\includegraphics[width=0.23\textwidth]{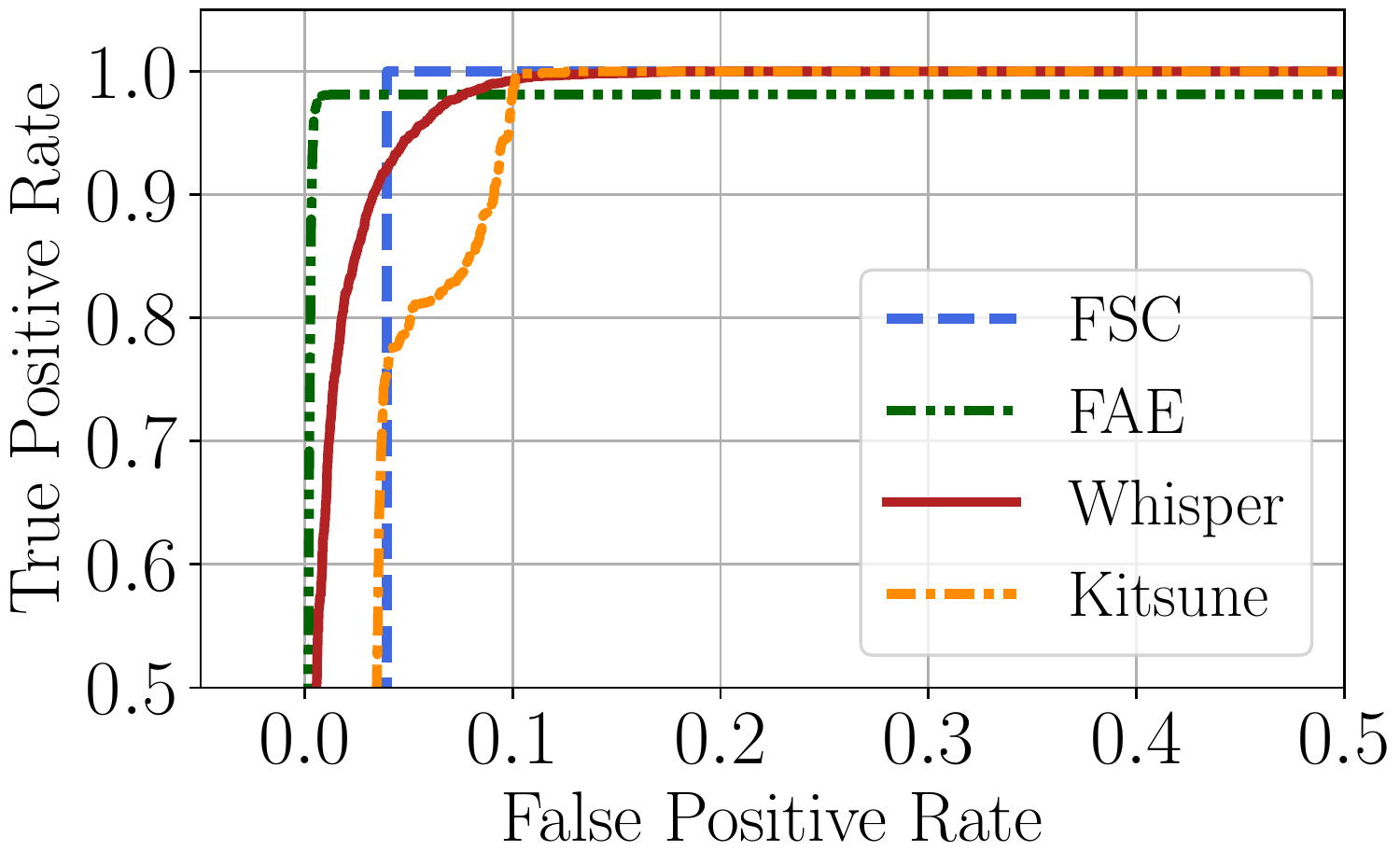}
	} 
	\hspace{-2mm}
	\subfigure [IPID Side-Channel (0.138 Mbps)]{ 
		\includegraphics[width=0.23\textwidth]{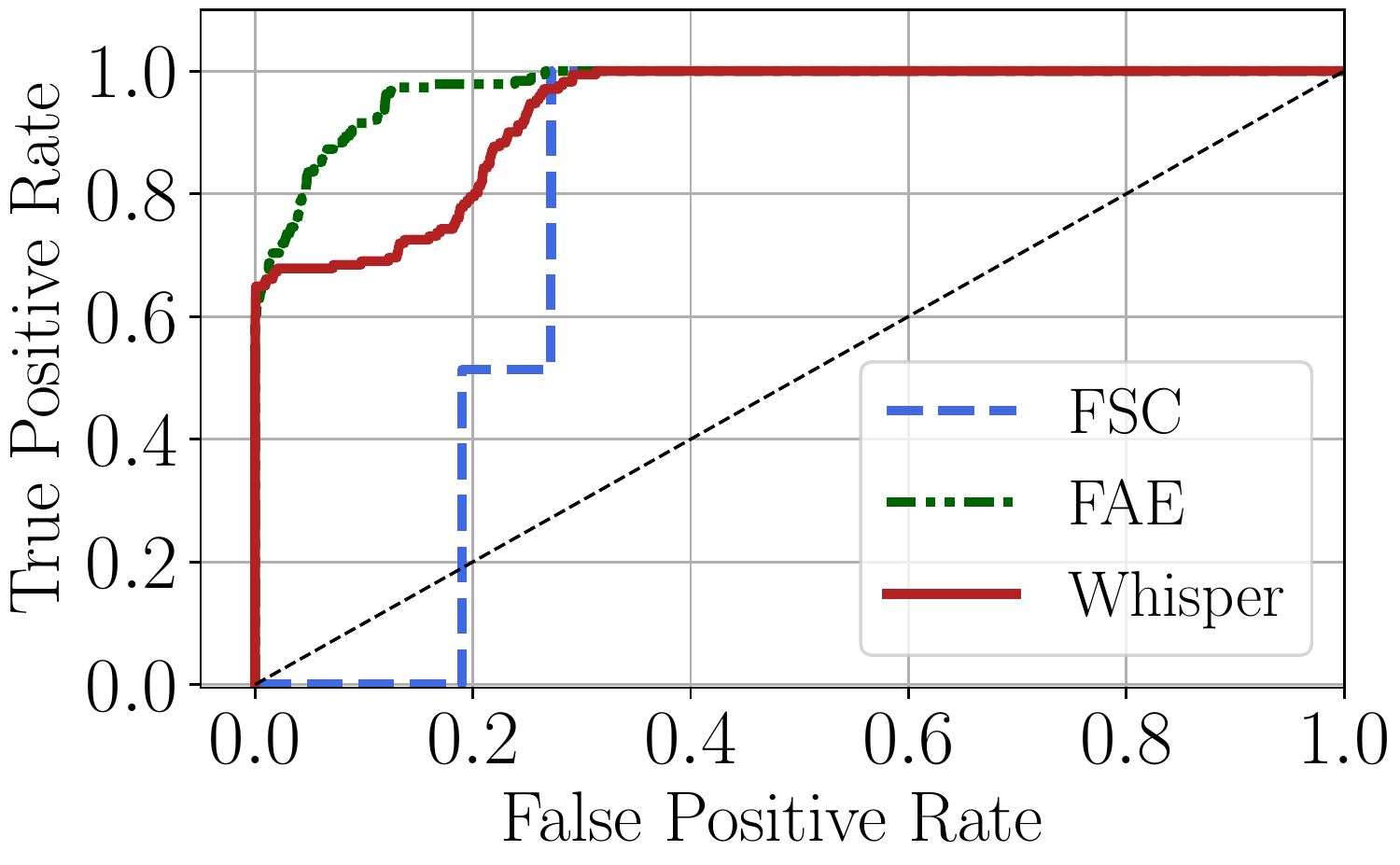}
	}
	\vspace{-4mm}
	\caption{ROC of high-rate attack: SYN DoS and low-rate attack: IPID side-channel attack.} 
	\label{graph:roc}
	\end{center}
	\vspace{-5mm}
\end{figure}

In this experiment, we evaluate the detection accuracy of different systems by measuring TPR, FPR, AUC, and EER. Table~\ref{table:comparsion} illustrates the results. We find that \NM\ can detect all 14 attacks with AUC ranging between 0.932 and 0.999 and EER within 0.201. Figure~\ref{graph:center} shows the scatter plots of clustering results. For simplicity, we select two datasets with 2,000 benign and 2,000 
malicious frequency domain features and choose two dimensions of the frequency domain features randomly. We observe that the malicious traffic has frequency domain features far from the clustering centers. We present the ROC curves of two datasets in Figure~\ref{graph:roc}. We find that, by leveraging the frequency domain features, detectors can detect low-rate malicious traffic in high throughput networks, e.g., \NM\ and FAE detect 138 Kbps IPID side-channel malicious traffic with 0.932 and 0.973 AUC under the 5.276 Gbps backbone network traffic, respectively. The increment of burst intervals in low-rate TCP DoS attacks causes 9.0\%, 7.0\%, 0.10\%, and 0.06\% AUC decrease for Kitsune, FSC, FAE, and \NM, respectively. Thus, compared with the packet-level and the traditional flow-level detection, burst intervals in the low-rate TCP DoS attacks have a negligible effect on the detection accuracy of \NM\ and FAE. However, FAE cannot effectively detect some sophisticated attacks, e.g., the ACK throttling side-channel attack and the TLS padding oracle attack, and only achieves only 39.09\% AUC of \NM. Note that, \NM\ accurately identifies 2.4 Gbps high-rate malicious flows among 4.8 Gbps traffic online.

Kitsune cannot effectively detect the side-channel attacks because it is unable to maintain enough states for the traffic. We find that Kitsune's offline processing speeds in the datasets are less than 4000 packets per second (PPS), and the expected time to complete the detection is more than 2 hours. The side-channel attacks trick Kitsune to maintain massive flow states by sending a larger number of probing packets. Different from using flow states to preserve the flow context information in Kitsune, \NM\ preserves the flow-level context information via the frequency domain analysis, which ensures the ability to detect such attacks.

We observe that, with the same ML algorithm, i.e., auto-encoder, the frequency domain features achieve higher accuracy (at most 15.72\% AUC improvements and 95.79\% EER improvements) than the state-of-the-art packet-level features and can detect more stealthy attacks. Under the five types of stealthy TCP attacks, Kitsune achieves 0.837 - 0.920 AUC and cannot detect the low-rate scanning of the side-channel attack. Moreover, compared with FSC, \NM\ achieves at most 65.26\% AUC improvements and 98.80\% EER improvements. Thus, we can conclude that the frequency domain features allow \NM\ to achieve higher detection accuracy and outperform the packet-level methods and the traditional flow-level methods.

\begin{figure}[t]
    \subfigcapskip=-2mm
    \begin{center}
	\subfigure[AUC comparison (higher is better)]{
		\includegraphics[width=0.47\textwidth]{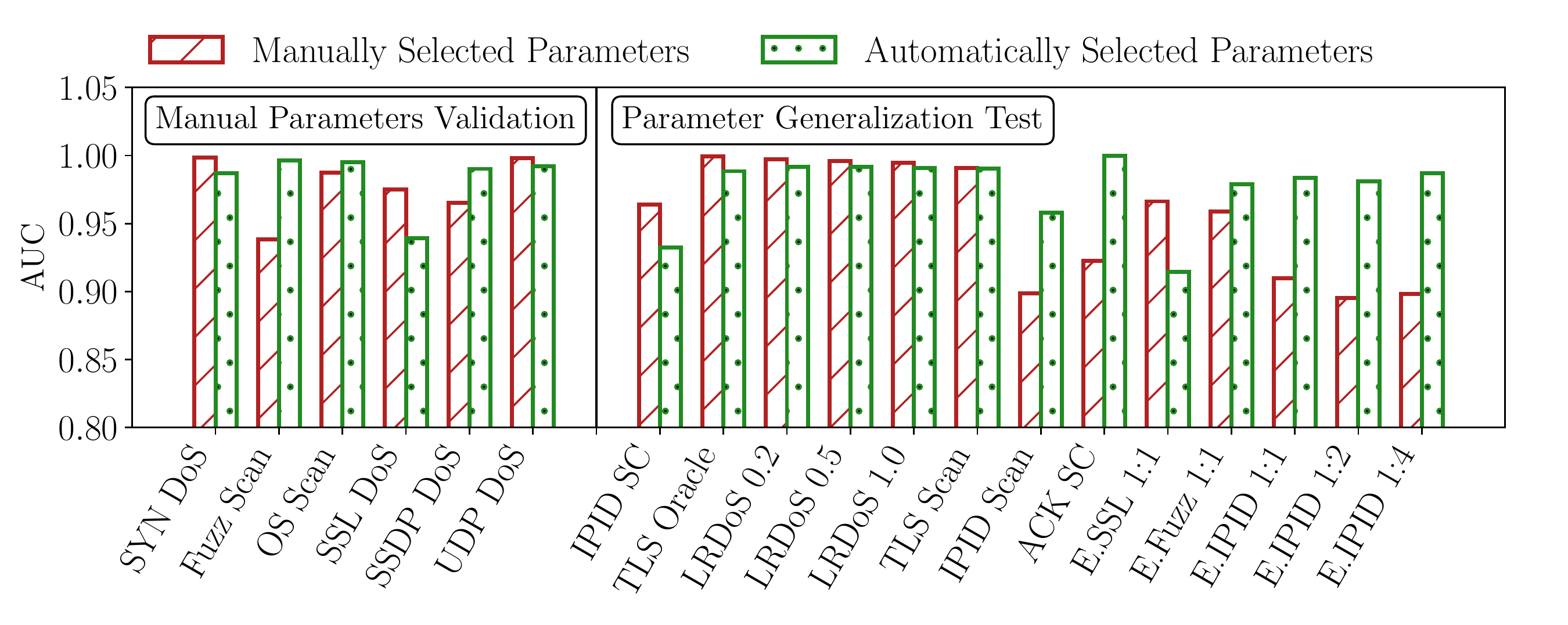}
	}
	\vspace{-4mm}
    \\
	\subfigure[EER comparison (lower is better)]{
		\includegraphics[width=0.47\textwidth]{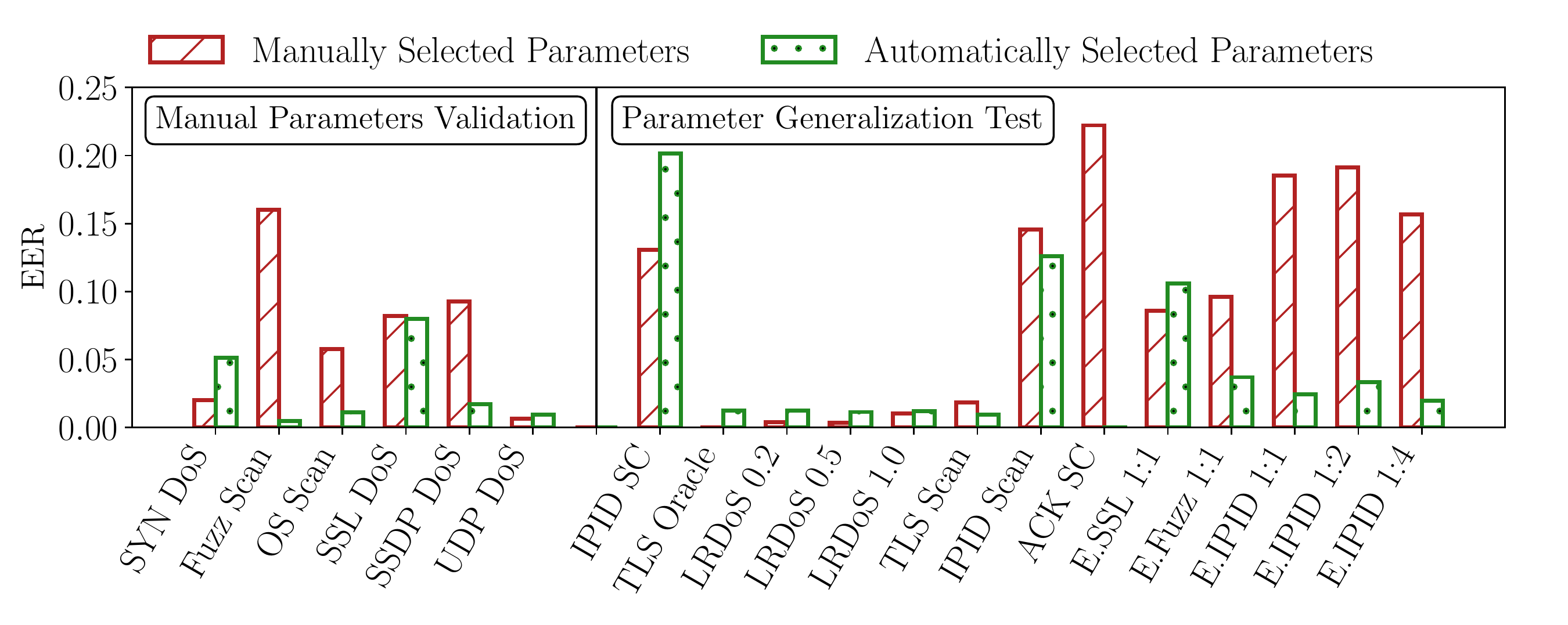}
	}
	\vspace{-4mm}
	\caption{Performance of the automatic parameter selection in comparison with manually selected parameters.} 
	\label{graph:param}
	\end{center}
\vspace{-5mm}
\end{figure}

Moreover, we study the impact of the automatic parameter selection on the detection accuracy. We manually set encoding vectors to compare the results with automatically selected parameters. We use six attacks as validation sets for the manually selected encoding vector, and use 13 attacks to test the generalization of the manually selected parameters. Figure~\ref{graph:param} shows the detection accuracy in terms of parameter settings. We observe that the automatic parameter selection module achieves 9.99\% AUC improvements and 99.55\% EER improvements compared with manual parameter selection.

\subsection{Robustness of Detection} \label{section:evaluation:robustness}
\begin{figure}[t]
    \subfigcapskip=-2mm
    \begin{center}
	\subfigure[Averaged AUC on different mix ratio (higher is better)]{
		\includegraphics[width=0.47\textwidth]{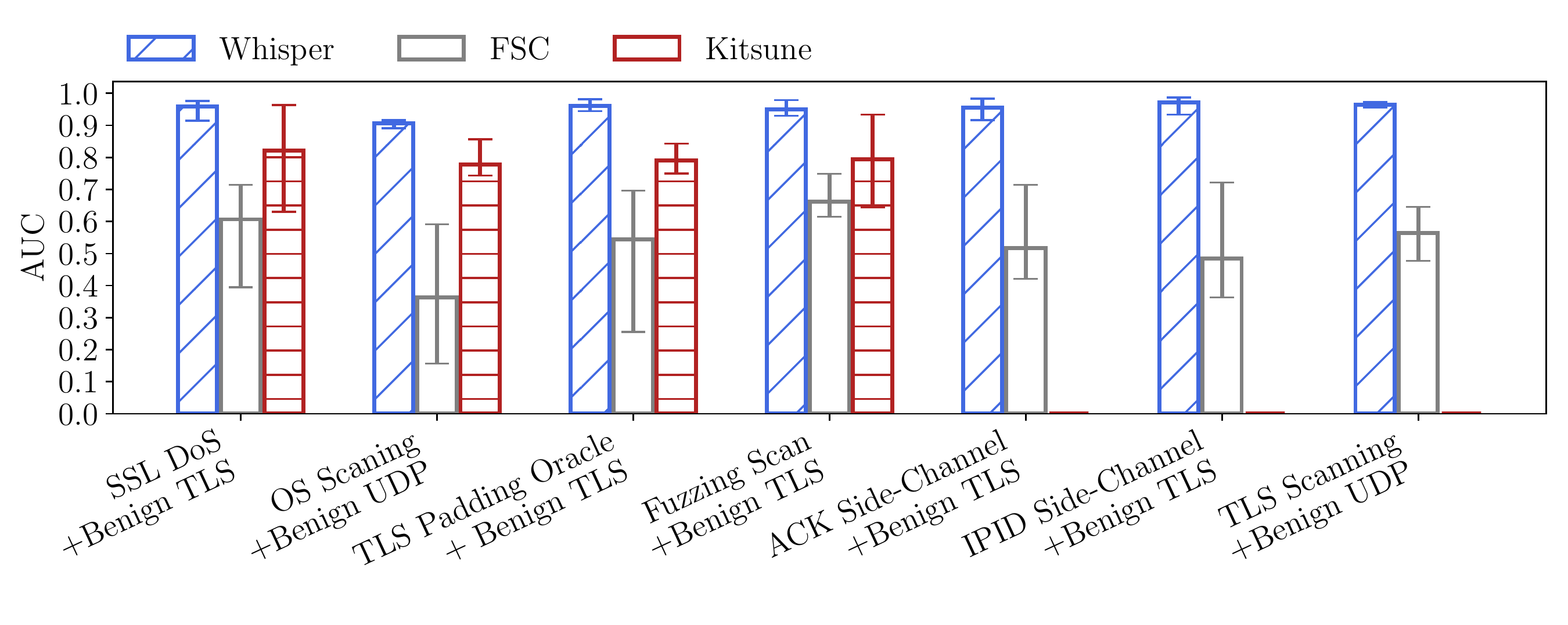}
	}
	\vspace{-4mm}
	\\
	\subfigure [Averaged EER on different mix ratio (lower is better)]{
		\includegraphics[width=0.47\textwidth]{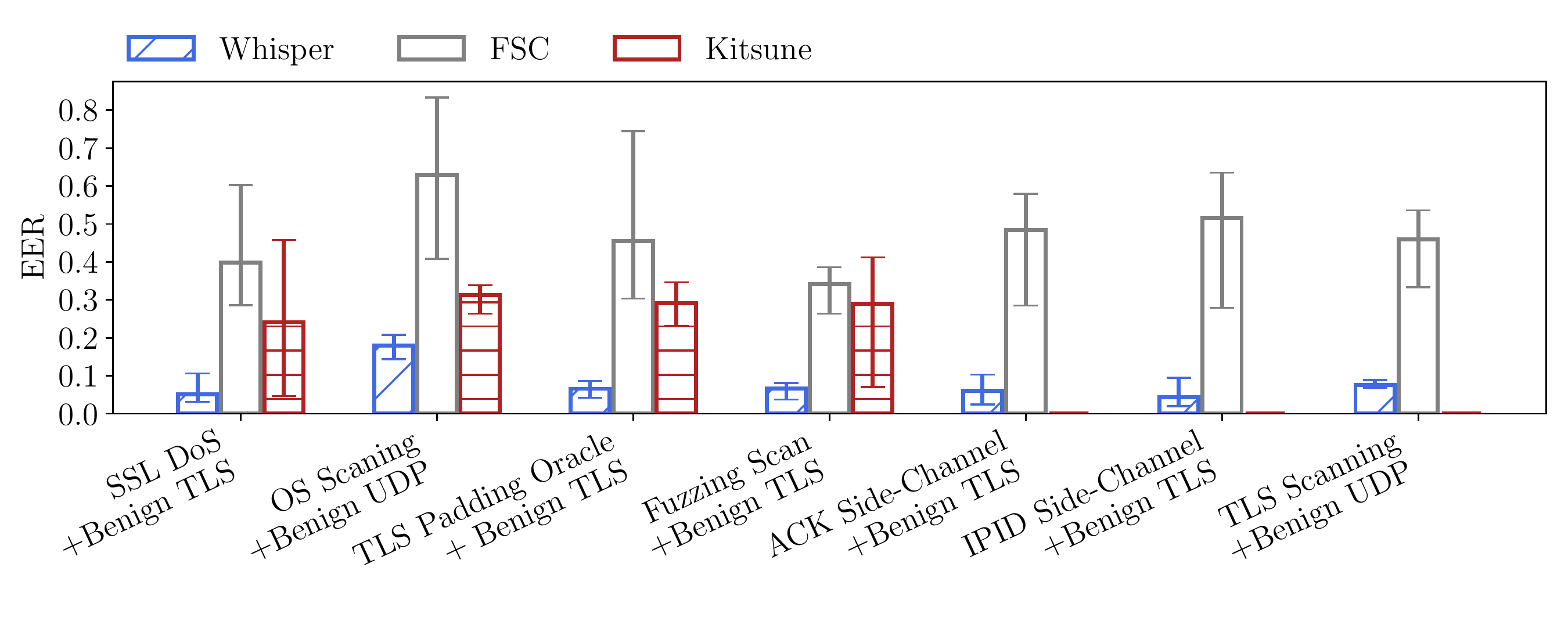}
	}
	\vspace{-4mm}
	\caption{Detection accuracy under attacks with various evading strategies.}
	\label{graph:robust-summary}
	\end{center}
\vspace{-5mm}
\end{figure}

In order to validate the robustness of \NM, we assume that attackers know the existence of malicious traffic detection. Attackers can construct evasion attacks, i.e., injecting various benign traffic, to evade the detection. In the experiments, for simplicity, we assume that attackers inject benign TLS traffic and UDP video traffic into the malicious traffic and disguise it as benign traffic for evasion. The reason why we use TLS and UDP video traffic is that it contributes to a high proportion of the benign traffic datasets, i.e., around 35\% and 13\%, respectively. Injecting the traffic can significantly interfere with traditional detection (see Figure~\ref{graph:robust-summary}). We select and replay 7 malicious traffic patterns and mix them into different ratio of benign traffic, i.e., the ratio of malicious traffic to the benign traffic ranging between 1:1 and 1:8. We do not inject the benign traffic with more ratio because the effectiveness of attacks is already low at the ratio of 1:8. We average the detection results with different ratio. Figure~\ref{graph:robust-summary} shows the averaged detection accuracy on different attacks. The detailed detection accuracy results can be found in Appendix~\ref{section:appendix:robust} (see Figure~\ref{graph:robustness}). We observe that the evasion attacks with high benign traffic mix ratio are prone to evade the detection. According to figure~\ref{graph:robust-summary}, we conclude that attackers cannot evade \NM\ by injecting benign traffic into the malicious traffic. However, the attackers evade other detection systems.

For instance, \NM\ has at most 10.46\% AUC decrease and 1.87 times EER increase under the evasion attacks. However, Kitsune has at most 35.4\% AUC decrease and 7.98 times EER increase. Similarly, attackers can effectively evade the detection of the traditional flow-level detection system, especially injecting more benign traffic with higher ratio. The evasion attacks, e.g, evasion OS scan and evasion TLS vulnerabilities scan, lead to at most 11.59 times EER increase under the flow-level methods (AUC $\le$ 0.5). Thus, we can conclude that the existing flow-level and packet-level detection systems are not robust to the evasion attacks. \NM\ has stable detection accuracy at different ratio, e.g., the averaged AUC decrease is bounded by 3.0\%, which is robust for evasion attacks. Moreover, We use other evading strategies to validate the robustness of \NM\ (see Appendix~\ref{section:appendix:robust}), e.g., injecting benign DNS, ICMP traffic and manipulating packet size and rate.

In summary, \NM\ can achieve robust detection because the used frequency domain features represent robust fine-grained sequential information of traffic. Malicious traffic disguised as benign traffic do not incur significant changes in the flow-level statistics. Thus, the features of the malicious traffic in the flow-level methods are the same to the benign flows. As a result, due to the invariant features, packet-level and traditional flow-level detection is unable to capture such attacks. For example, the packet-level methods (e.g., Kitsune) use the statistics as the context information. However, the sequential information of the malicious traffic extracted by \NM\ are significantly different from the benign traffic. Thus, to our best knowledge, \NM\ is the first machine learning based method that achieves robust detection under evasion attacks. 

\subsection{Detection Latency and Throughput}\label{section:evaluation:runtime}
\begin{figure*}[h]
    \subfigcapskip=-2mm
    \begin{center}
	\subfigure[Overall latency (processing and queuing)]{
		\includegraphics[width=0.30\textwidth]{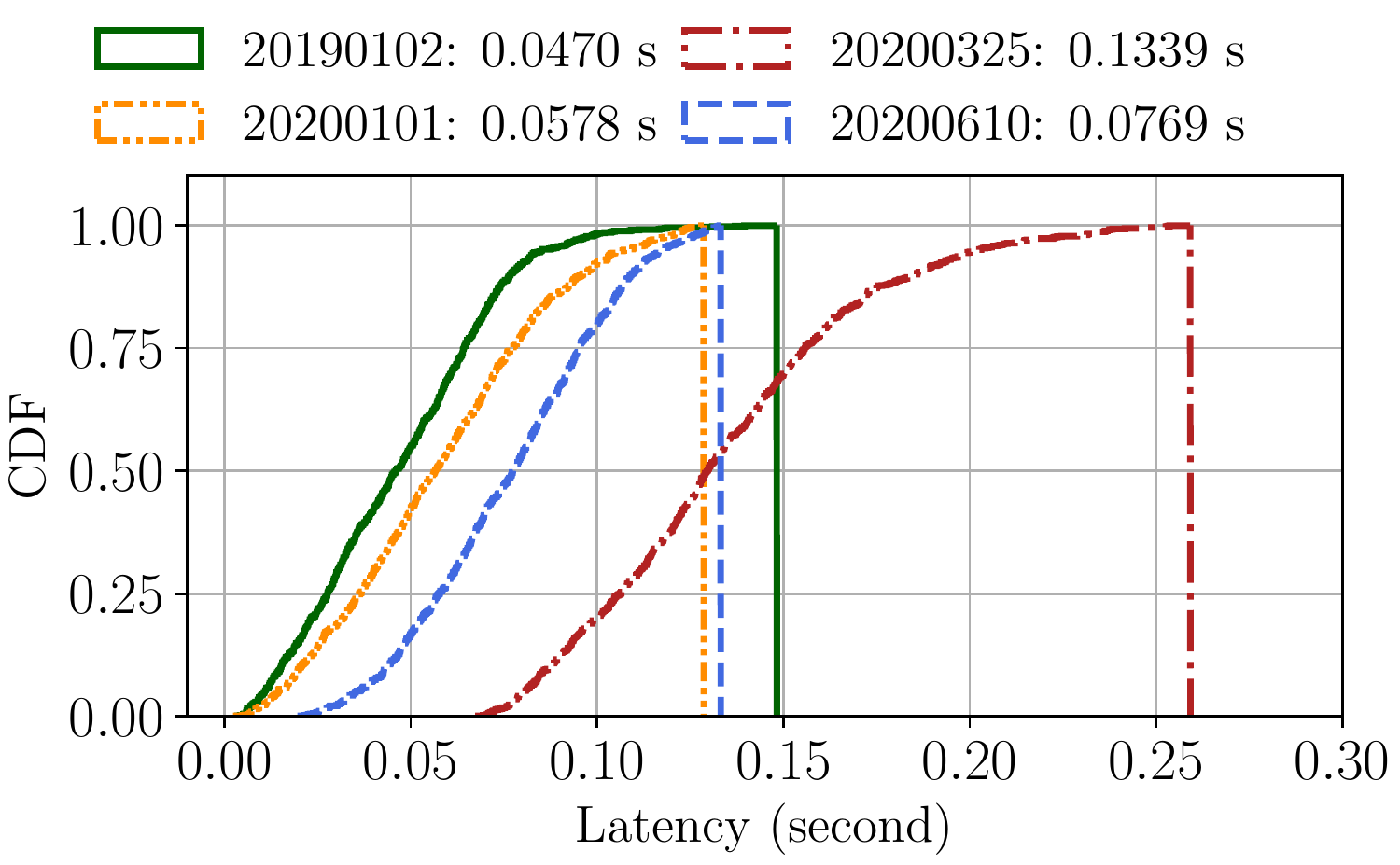}
		\label{graph:delay:overall}
	} 
	\hspace{1mm}
	\subfigure [Pure processing latency]{ 
		\includegraphics[width=0.30\textwidth]{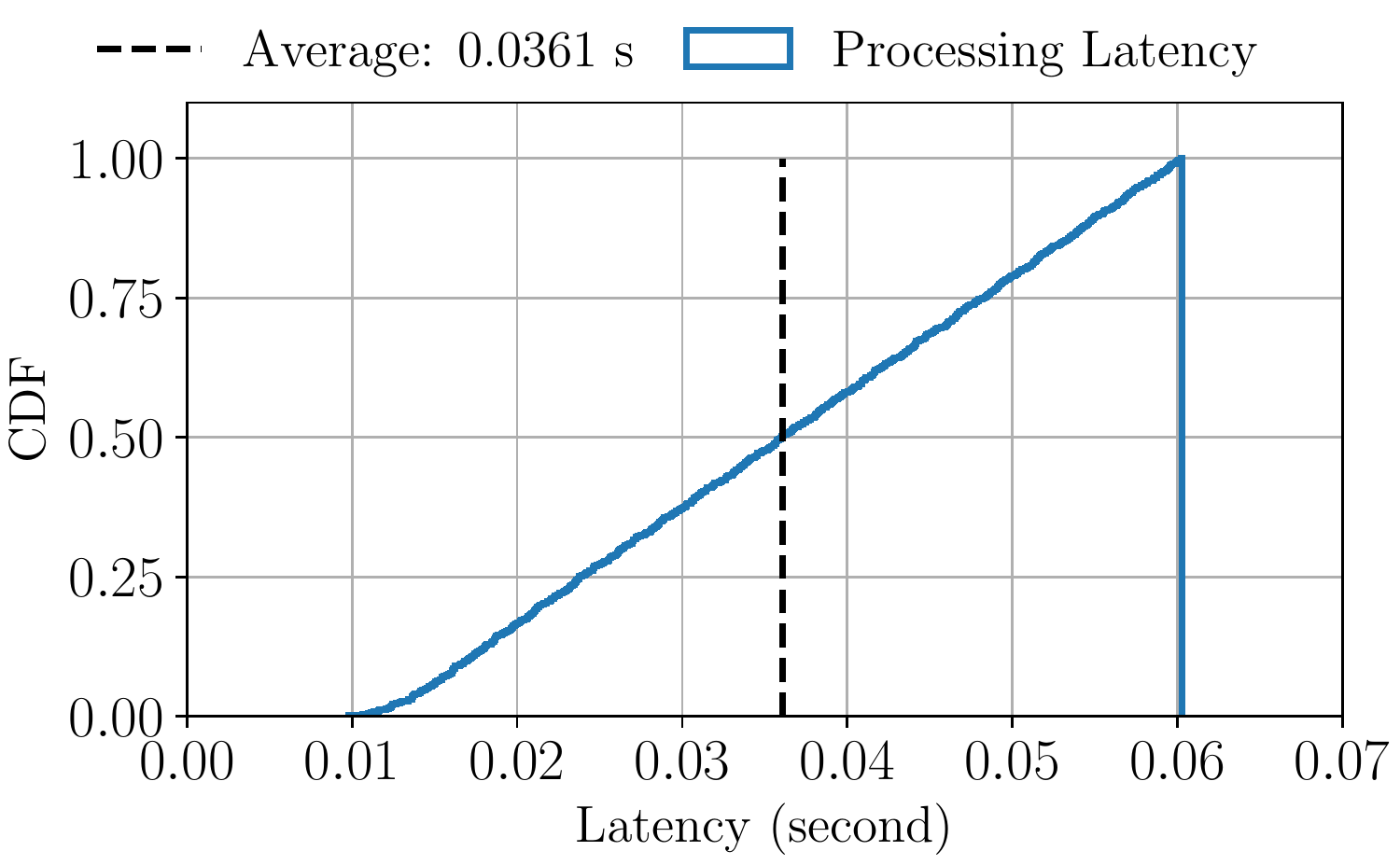}
		\label{graph:delay:pure}
	}
	\hspace{1mm}
	\subfigure [Processing latency of different steps]{ 
		\includegraphics[width=0.30\textwidth]{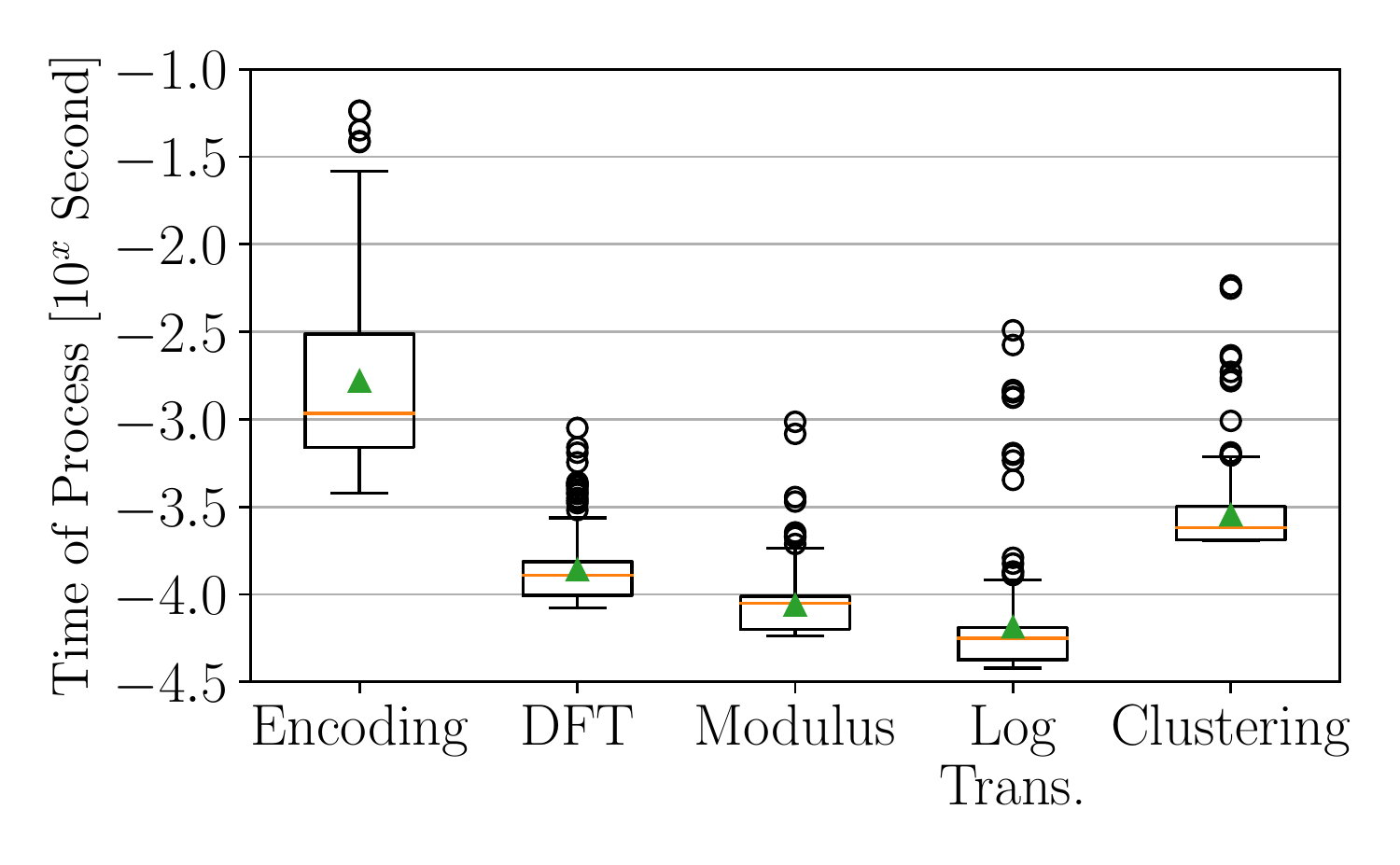}
		\label{graph:delay:setp}
	}
	\vspace{-4mm}
	\caption{Detection latency of \NM. We present the CDF of overall latency in (a), the CDF of pure processing latency in (b), the box plot of latency in different steps in (c).}
	\label{graph:delay}
	\end{center}
\vspace{-6mm}
\end{figure*}

\noindent \textbf{Detection Latency.}
To measure the latency, we replay the backbone network traffic datasets with different traffic rates (see Table~\ref{table:dataset}). For simplicity, we use the low-rate TCP DoS attack with a 0.5s burst interval as a typical attack and measure the overall detection latency, i.e., the time interval between sending the first malicious packet and detecting the traffic. The overall detection latency includes the transmitting latency, the queuing latency, and the processing latency. The cumulative distribution function (CDF) of the overall detection latency is shown in Figure~\ref{graph:delay:overall}. With four datasets, we find that the detection latency of \NM\ is between 0.047 and  0.133 second, which shows that \NM\ achieves realtime detection in high throughput networks. In order to accurately measure the processing latency incurred by \NM, we replay the low-rate TCP DoS dataset with a 0.5s burst interval to construct a light load network scenario and measure the execution time of the four modules in \NM. The CDF of the processing latency is shown in Figure~\ref{graph:delay:pure}. We observe that the processing latency of \NM\ exhibits uniform distribution because most of the latency is incurred by polling per-packet features from the packet parser module in the light load situation. Thus, we can conclude that the averaged processing latency incurred by \NM\ is only 0.0361 second, and the queuing latency raised by \NM\ is the majority. 

\begin{figure*}[h]
    \subfigcapskip=-3mm
    \begin{center}
	\subfigure[\NMM]{
		\includegraphics[width=0.30\textwidth]{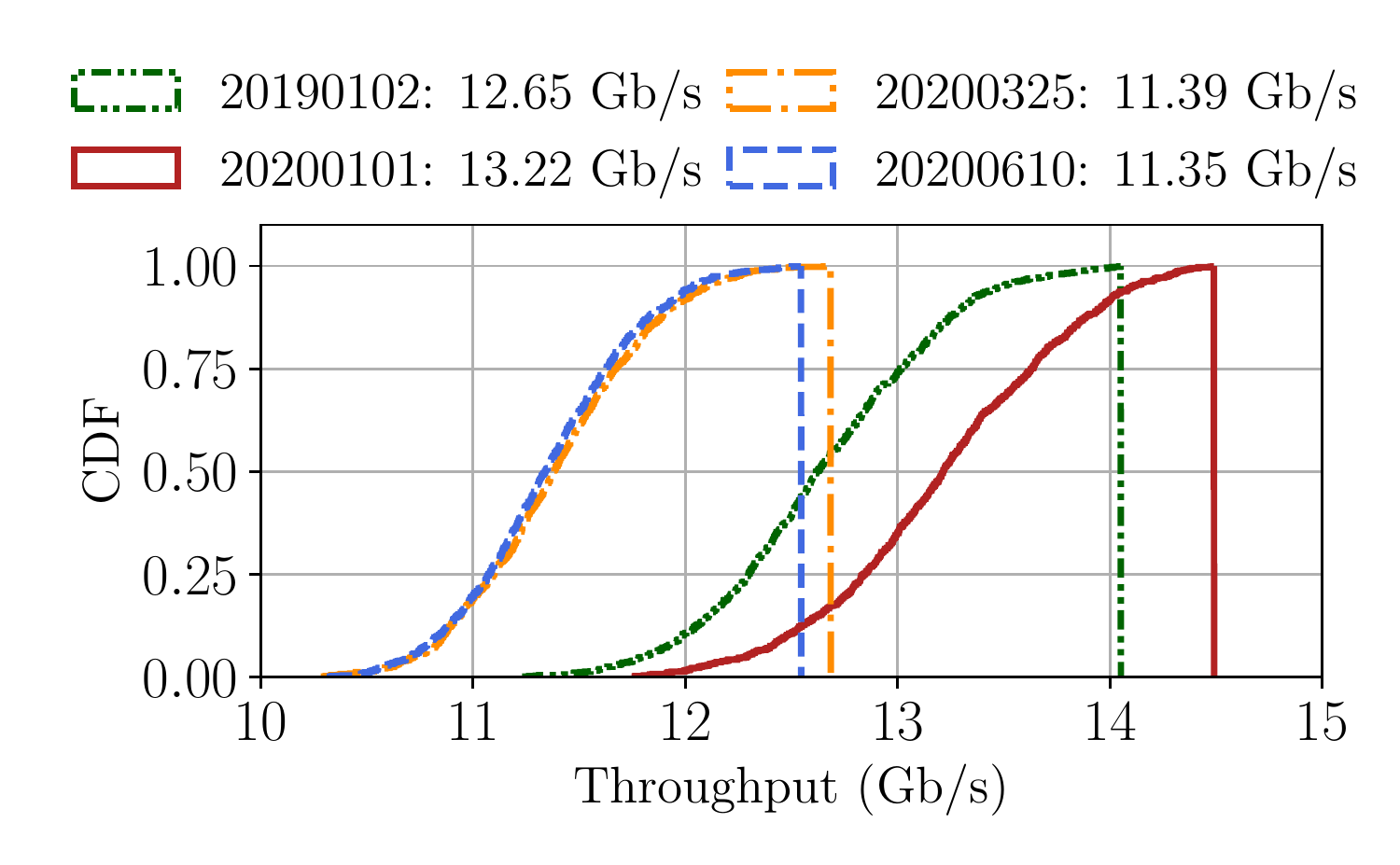}
	}
	\hspace{1mm}
	\subfigure [FAE]{
		\includegraphics[width=0.30\textwidth]{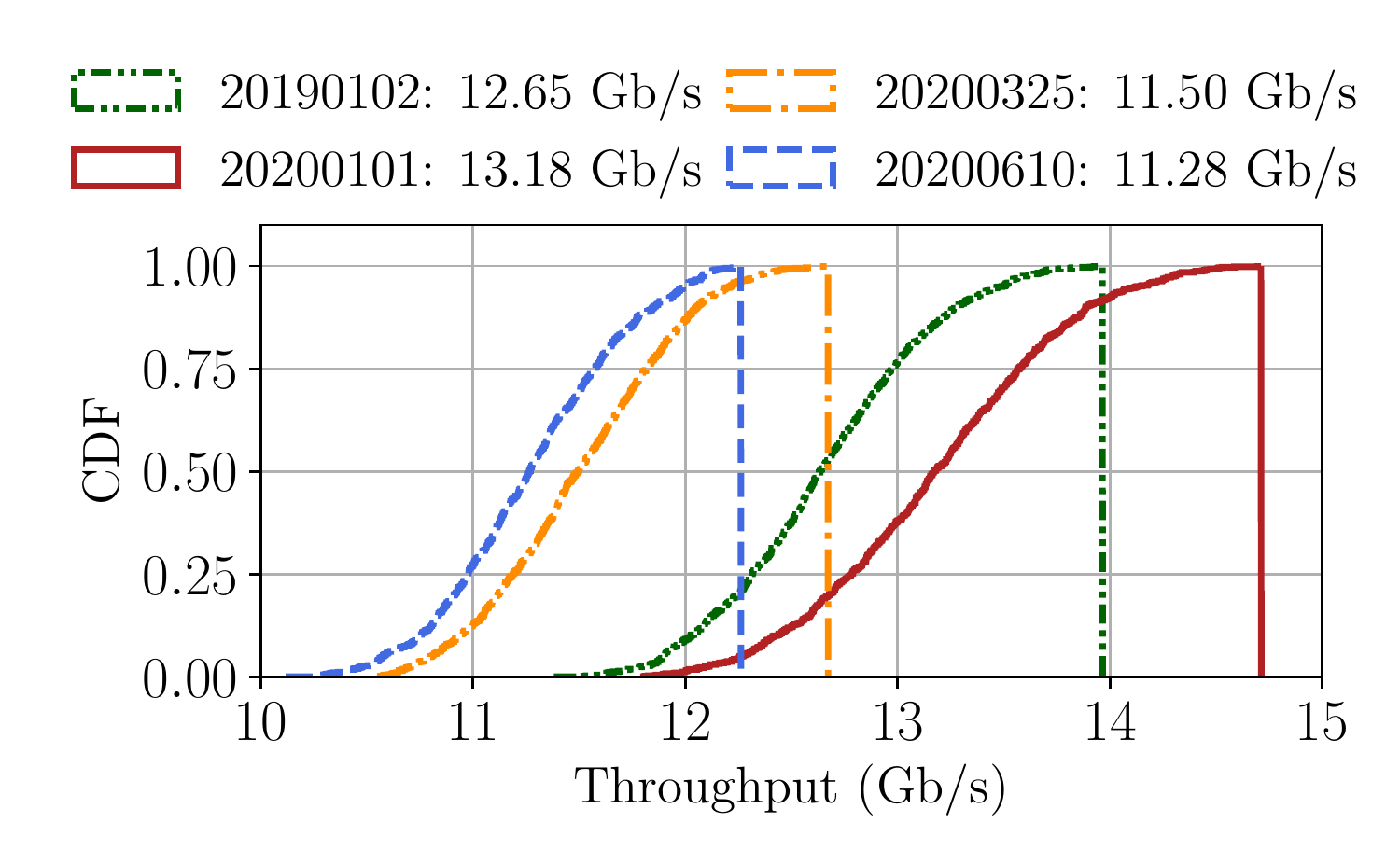}
	}
	\hspace{1mm}
	\subfigure [Kitsune]{
		\includegraphics[width=0.30\textwidth]{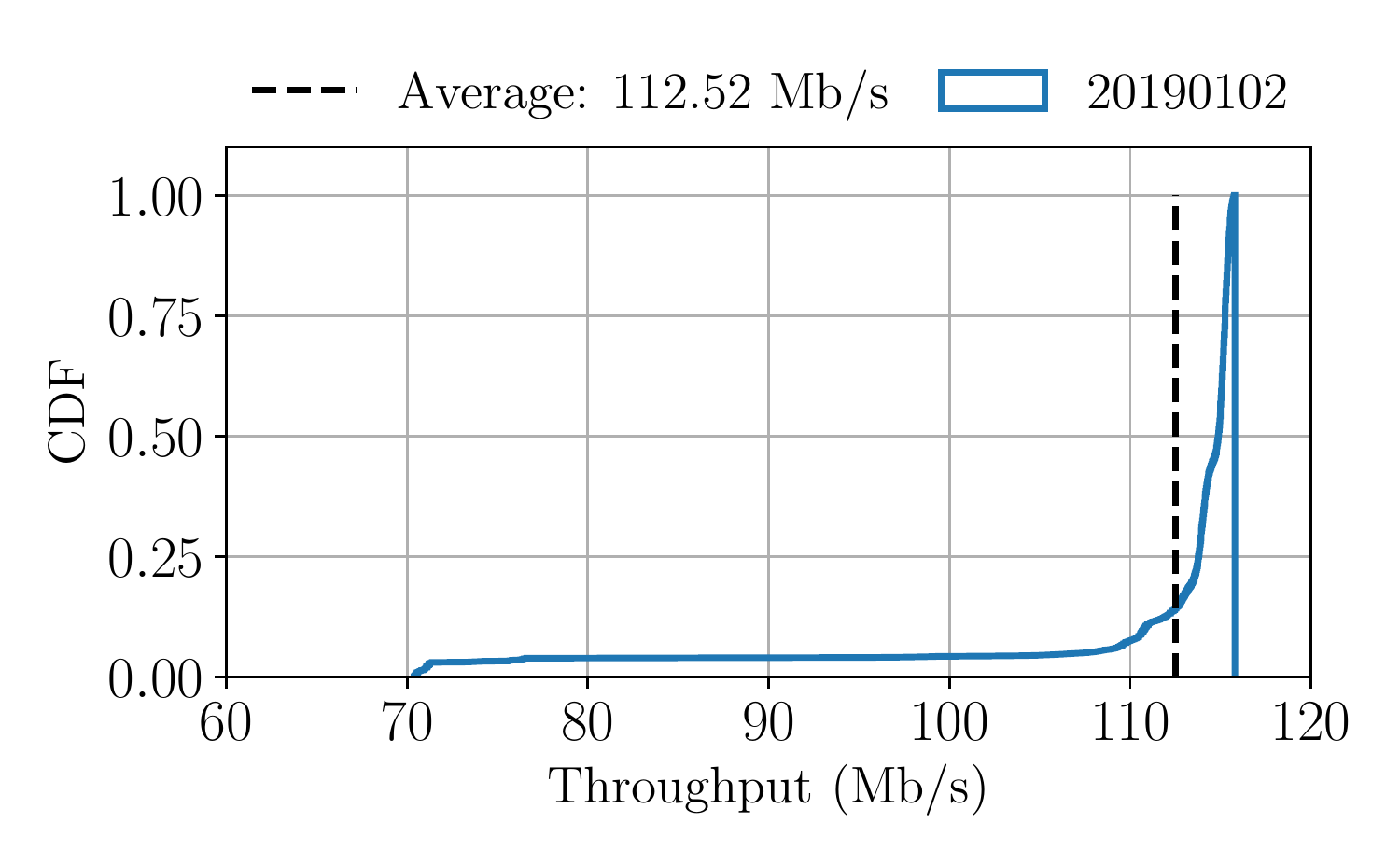}
	}
	\vspace{-4mm}
	\caption{CDF and the average number of throughput: \NM, FAE, and Kitsune.}
	\label{graph:throughput}
	\end{center}
\vspace{-3mm}
\end{figure*}

We also analyze the latency raised by each step of \NM\ in Figure~\ref{graph:delay:setp}. We see that the measured latency in each step is consistent with the computational complexity analysis in Section~\ref{section:analysis:scale}. The DFT, Modulus Calculation, and Log Transformation have similar computational complexity and incur similar processing latency. The most latency is raised from the packet encoding (i.e., $5.20 \times 10^{-3}$ second on average). The statistical clustering module has averaged processing latency of $1.30 \times 10^{-4}$ second, which is significantly lower than the packet encoding. We find that most of the latency is incurred by the packet parsing module and the memory copy for parsing per-packet features incurs the most latency.

\noindent \textbf{Throughput.} We replay four MAWI~\cite{WIDE} backbone network traffic datasets with the physical limit bandwidth of laser ports (20 Gbps) to measure the throughput. We measure the throughput of \NM\ and FAE and validate that detection accuracy does not decrease when reaching the maximum throughput. We run Kitsune with the same hardware as \NM\ and measure the offline processing speed, i.e., we ignore the packet parsing overhead in the online processing of Kitsune, because it cannot handle high speed traffic. The CDF of the throughput is shown in Figure~\ref{graph:throughput}. We find that \NM\ achieves 11.35 Gbps to 13.22 Gbps average throughput, while Kitsune achieves 112.52 Mbps. \NM\ achieves high throughput because it significantly reduces the processing overhead of the machine learning. FAE achieves the averaged throughput ranging between 11.28 Gbps and 13.18 Gbps, which is similar to \NM. Note that, FAE uses a similar auto-encoder model in Kitsune and achieves 100 times higher throughput (though it has limited detection ability). We conclude that the frequency domain features used in \NM\ enable higher throughput than the packet-level methods. In summary, \NM\ and FAE achieve the most throughput, around 1.27 $\times10^6$ PPS, compared with other detection systems.

\section{Related Work} \label{section:related}
\textbf{Machine Learning based NIDS.} Machine learning based Network Intrusion Detection Systems (NIDSes) can achieve higher detection accuracy than the traditional signature based NIDSes~\cite{Snort, Suricata, USEC12-Chimera, CCS18-vNIDS}. In particular, compared with the signature based NIDSes, they can detect zero-day attacks that have not been uncovered~\cite{ACM-CS-16-survey, C&S09-survey}. For example, Antonakakis~\etal~\cite{USEC12-Botnet-Traffic},  Nelms~\etal~\cite{USEC15-WebWitness}, and Invernizzi~\etal~\cite{NDSS14-Nazca} detect malware traffic by using statistical machine learning approaches. Moreover, the specialized features of botnets have been used in botnet traffic detection~\cite{NDSS16-who, ACSAC18-Lobo, S&P12-Evilseed, S&P20-Wang}. Different from these methods, \NM\ detects various attack traffic including botnet traffic online. Bartos~\etal~\cite{USEC16-Variants} developed an invariant of statistical features based detection via matrix transformations, which is not scalable in large scale detection. Mirsky~\etal~\cite{NDSS18-Kitsune} proposed Kitsune that leveraged lightweight deep neural networks, i.e., auto-encoders, to reduce the processing overhead. \NM\ uses packet encoding and DFT to compress the original per-packet features for reducing feature redundancy. The compressed frequency domain features allow the machine learning to be readily deployable for high performance detection.

\noindent \textbf{Traffic Classification.} Machine learning algorithms are widely used in traffic classification~\cite{TIFS17-Meng, IEEE-Netw-Meng, NDSS20-DNS-FP, NDSS20-instantaneous-FP, NDSS20-mobile-app-FP, CCS17-Scalable-TA, TIFS21-DAPP, TIFS21-length-only-FP, CaoRAID19}. For example, web fingerprinting aims to invalidate the Tor anonymous services and infer the website that users are visiting by using the features of TLS encrypted traffic~\cite{NDSS18-Auto-FP,XuACSAC18,YinTDSC21}. Similar to Web fingerprinting, Ede~\etal~\cite{NDSS20-mobile-app-FP} used semi-supervised learning to fingerprint mobile applications. Siby~\etal~\cite{NDSS20-DNS-FP} applied traffic analysis to classify encrypted DNS traffic and infer the activities of users. Bahramali~\etal~\cite{NDSS20-instantaneous-FP} analyzed the features of various realtime communication applications. Nasr~\etal~\cite{CCS17-Scalable-TA} compressed the statistical features of traffic, which achieved large scale traffic analysis. Zhang~\etal~\cite{NDSS19-defence-TA} proposed a countermeasure against traffic analysis via adversarial examples. Although traffic classification achieves a different goal from malicious traffic detection and cannot be used in traffic detection, the extracted traffic features in \NM, i.e., the frequency domain features, can be applied to perform traffic classifications.

\noindent \textbf{Anomaly Detection with Data Augmentation.} Data augmentation is recently developed efficiently model training for anomaly detection~\cite{S&P20-Wang,USEC18-A4NT,USEC19-StackOver}. For example, Jan~\etal~\cite{S&P20-Wang} leveraged Generative Adversarial Network (GAN) to generate labeled datasets for botnet detection. Shetty~\etal~\cite{USEC18-A4NT} generated paired data by using GAN to train a seq2seq model that aims to invalidate the anonymity of text. Fischer~\etal~\cite{USEC19-StackOver} solved the dataset scalability problem to detect vulnerable code via Siamese Networks. In \NM, we leverage the frequency domain features for efficient anomaly detection. 

\noindent \textbf{Throttling Malicious Traffic.} IP blacklists have been widely used to throttle malicious traffic~\cite{USEC19-Platforms, USEC19-Reading}. For instance, Ramanathan~\etal~\cite{NDSS20-BLAG} proposed an IP blacklist aggregation method to locate attackers. Moreover, programmable data planes~\cite{NDSS20-Poseidon,TON21-Poseidon,USEC20-NetWarden,TIFS18-Qi,USEC21-Ripple} have been recently leveraged to throttle various attack traffic, e.g., throttling different types of DoS flows and covert channels. All these defenses are orthogonal to our \NM.

\section{Conclusion} \label{section:conclusion}
In this paper, we develop \NM, a realtime malicious traffic detection system that utilizes sequential information of traffic via frequency domain analysis to enable robust attack detection. The frequency domain features with bounded information loss allow \NM\ to achieve both high detection accuracy and high detection throughput. In particular, fine-grained frequency domain features represent the ordering information of packet sequences, which ensures robust detection and prevents attackers from evading detection. In order to extract the frequency domain features, \NM\ encodes per-packet feature sequences as vectors and uses DFT to extract sequential information of traffic in the perspective of frequency domain, which enables efficient attack detection by utilizing a lightweight clustering algorithm. We prove that the frequency domain features have bounded information loss which is a prerequisite of accuracy and robustness. Extensive experiments show that \NM\ can effectively detect various attacks in high throughput networks. It achieves 0.999 AUC accuracy within 0.06 second and around 13.22 Gbps throughput. Especially, even under sophisticated evasion attacks, \NM\ can still detect malicious flows with high AUC ranging between 0.891 and 0.983.\\[2mm]

\begin{acks}
We thank the anonymous reviewers for their insightful comments. This work was in part supported by the National Key R\&D Program of China with No.2018YFB0803405, China National Funds for Distinguished Young Scientists with No.61825204, National Natural Science Foundation of China with No.61932016 and No.62132011, Beijing Outstanding Young Scientist Program with No.BJJWZYJH01201 910003011, BNRist with No.BNR2019RC01011. Ke Xu is the corresponding author of this paper.
\end{acks}
\newpage

\bibliographystyle{acmref}
\bibliography{input}
\newpage

\appendix
\section*{Appendix} \label{section:appendix}
  
\section{Proof of Theorem 1} \label{section:appendix:1}
$\mathcal{H}_{\mathrm{packet}}$ denotes the overall differential entropy of the sampling sequence $\vec{s}$, i.e., the sum of the differential entropy of each random variable in $\vec{s}$: ($K = \sqrt{2\pi e}$)
\begin{small}
\begin{align*}
    \mathcal{H}_{\mathrm{packet}} &= - \sum^{N}_{i=1} \int_{-\infty}^{+\infty} p_i(s) \ln p_i(s) \mathrm{d}s \\
                                  &= \sum^{N}_{i=1} \ln \sigma(i) K \\
                                  &= \ln K^N \prod^{N}_{i=1} \sigma(i).
\end{align*}
\end{small}

\noindent We assume that the statistical feature extraction function $\mathsf{f}$ calculates the minimum of $\vec{s}$ to acquire the flow-level features. $I_{\mathrm{min}}$ denotes the index of the sample with the minimum value. The differential entropy of the feature is $\mathcal{H}_{\mathrm{flow-min}}$ that equals to the entropy of the random variable with the minimum value:
\begin{small}
\begin{align*}
    I_{\mathrm{min}} &= \mathop{\arg\min }_{i} s_i ,
\end{align*}
\begin{align*}
    \mathcal{H}_{\mathrm{flow-min}} &= - \int_{-\infty}^{+\infty} p_{I_{\mathrm{min}}}(s) \ln p_{I_{\mathrm{min}}}(s) \mathrm{d}s \\
                                    &= \ln K \sigma(I_{\mathrm{min}}).
\end{align*}
\end{small}

\noindent $\Delta\mathcal{H}_{\mathrm{flow-min}}$ denotes the differential entropy loss of the minimum feature, i.e., the difference between the overall differential entropy and the differential entropy of the minimum feature:
\begin{small}
\begin{align*}
    \Delta\mathcal{H}_{\mathrm{flow-min}} &= \mathcal{H}_{\mathrm{packet}} - \mathcal{H}_{\mathrm{flow-min}} \\
                                          &= \ln K^{N-1} \prod_{i \ne I_{\mathrm{min}}} \sigma(i).
\end{align*}
\end{small}

\noindent We focus on the expectation of the loss, and leverage Jensen inequation to get the lower bound of the information loss:
\begin{small}
\begin{align*}
    \mathrm{E}[\Delta\mathcal{H}_{\mathrm{flow-min}}] &\ge \ln K^{N - 1} E[\sigma^{N - 1}] \\
                                             &\ge (N - 1) \ln K \mathrm{E}[\sigma].
\end{align*}
\end{small}

\noindent We conduct the same proof procedure for the features that calculate the maximum of the per-packet feature sequence and complete the proof of Theorem 1.

\section{Proof of Theorem 2 and Theorem 3} \label{section:appendix:2-3}
We consider the situation that a flow-level feature extraction method calculates the average number of sampled per-packet features. We denote the average of $\vec{s}$ as a random variable $f_m$ that obeys a Gaussian distribution:
\begin{small}
\begin{align*}
    f_m \sim \mathcal{N}(\frac{1}{N} \sum^{N}_{i=1} u(i), \frac{1}{N^2} \sum^{N}_{i=1} \sigma^2(i)).
\end{align*}
\end{small}

\noindent $p_m$ denotes the probability density function (PDF) of $f_m$. We use $\mathcal{H}_{\mathrm{flow-avg}}$ and $\Delta\mathcal{H}_{\mathrm{flow-avg}}$ to indicate the differential entropy of the average feature and the information loss, respectively:
\begin{small}
\begin{align*}
    \mathcal{H}_{\mathrm{flow-avg}} &= - \int_{-\infty}^{+\infty} p_{m}(s) \ln p_{m}(s) \mathrm{d}s \\
                                     &= \ln \frac{K}{N} \sqrt{\sum^{N}_{i=1} \sigma^2(i)},
\end{align*}
\begin{align*}
    \Delta\mathcal{H}_{\mathrm{flow-avg}} &= \mathcal{H}_{\mathrm{packet}} - \mathcal{H}_{\mathrm{flow-avg}} \\
                                           &= \ln N K^{N-1} \frac{\prod_{i=1}^{N} \sigma(i)}{\sqrt{\sum^{N}_{i=1} \sigma^2(i)}}.
\end{align*}
\end{small}

\noindent To get the upper bound, we use $Q$ to indicate the square mean of the variances of $\vec{s}$. According to the inequality of arithmetic and geometric means, the geometric mean is not bigger than the square mean. We get the upper bound of the differential entropy loss:
\begin{small}
\begin{align*}
    \Delta\mathcal{H}_{\mathrm{flow-avg}} &\le \ln N K^{N-1} \frac{Q^N}{\sqrt{\sum^{N}_{i=1} \sigma^2(i)}} \\
                                           &\le \ln \sqrt{N} K^{N - 1} Q^{N - 1}.
\end{align*}
\end{small}If and only if $\sigma(i)$ is a constant, the information loss $\Delta\mathcal{H}_{\mathrm{flow-avg}}$ reaches its maximum. We use $\sigma_{\mathrm{max}}$ to indicate the maximum of the variances of $\vec{s}$, and get the lower bound of the information loss by leveraging the non-negative differential entropy assumption:
\begin{small}
\begin{align*}
    \sigma_{\mathrm{max}} = \mathsf{max}(\sigma(i)) \quad (1 \le i \le N),
\end{align*}
\begin{align*}
    \Delta\mathcal{H}_{\mathrm{flow-avg}} &\ge \ln \sqrt{N} K^{N-1} \frac{\prod_{i=1}^{N} \sigma(i)}{\sigma_{\mathrm{max}}} \\
                                            &\ge \ln \sqrt{N} \qquad (K\sigma(i) \ge 1).
\end{align*}
\end{small}The equality holds iff. $\sigma(i)=\frac{1}{K}$. When the equality holds, the upper bound equals the lower bound. Here we complete the proof of Theorem 3. Similar to the proof of Theorem 1, we apply Jensen inequation to get $\Delta\mathcal{H}_{\mathrm{flow-avg}}$ and prove Theorem 2.

\section{Proof of Theorem 4} \label{section:appendix:4}
We consider the situation that a flow-level feature extraction method calculates the variance of the sampling sequence to extract the features of traffic. Random variable $V$ denotes the variance of $\vec{s}$:
\begin{small}
\begin{align*}
    V = \frac{\sum_{i=1}^N (s_i - u)^2}{N},\quad u = \frac{\sum_{i=1}^N s_i}{N}.
\end{align*}
\end{small}

\noindent The random variable $V$ obeys general Chi-square distribution. We assume that the Gaussian process $\mathcal{S}$ is strictly stationary with zero mean, i.e., $u(i)=0$ and $\sigma(i)=\sigma$. We present an estimate of differential entropy loss when $N$ is large enough:
\begin{small}
\begin{align*}
    V=\frac{\sum_{i=1}^{N}s_i^2}{N} = \frac{\sigma^2}{N} \sum_{i=1}^{N}(\frac{s_i}{\sigma})^2, \quad \sum_{i=1}^{N}(\frac{s_i}{\sigma})^2 \sim \chi^2(N).
\end{align*}
\end{small}$\mathcal{H}_{\mathrm{flow-var}}$ denotes the differential entropy of the variance feature:
\begin{small}
\begin{align*}
    \mathcal{H}_{\mathrm{flow-var}} &= \mathcal{H}[V] = \mathcal{H}[\frac{\sum_{i=1}^{N}s_i^2}{N}]\\
                                    &= \ln \frac{\sigma^2}{N} + \mathcal{H}[\sum_{i=1}^{N}(\frac{s_i}{\sigma})^2]\\
                                    &= \ln \frac{\sigma^2}{N} + \ln 2 \Gamma(\frac{N}{2}) + (1 - \frac{N}{2}) \psi(\frac{N}{2}) + \frac{N}{2},
\end{align*}
\end{small}where $\Gamma$ is Gamma function and $\psi$ is Digamma function. When $N$ is large enough we take the even number that is closest to $N$ to approach the information loss: ($\gamma$ is Euler–Mascheroni constant)

\begin{small}
\begin{equation*}
    \begin{cases}
    \psi(x) &= \frac{\Gamma^{'}(x)}{\Gamma(x)} \\
    \Gamma(x) &= (x-1)! \\
    \Gamma^{'}(x) &= (x-1)! (-\gamma + \sum_{k=1}^{x-1}\frac{1}{k})
\end{cases}
\end{equation*}

\begin{equation*}
    \Rightarrow\mathcal{H}_{\mathrm{flow-var}} = \ln \frac{\sigma^2}{N} + \ln 2 (\frac{N}{2})! - \frac{N}{2} (-\gamma + \sum_{k=1}^{\frac{N}{2}}\frac{1}{k}) + \frac{N}{2}.
\end{equation*}
\end{small}

\noindent Then we approach the Harmonic series as follows,

\begin{small}
\begin{align*}
    \sum_{k=1}^{N} \frac{1}{k} &\approx \ln N + \gamma, \\
    \Rightarrow\mathcal{H}_{\mathrm{flow-var}} &= \ln \frac{\sigma^2}{N} + \ln 2 (\frac{N}{2})! - \frac{N}{2}\ln\frac{N}{2} + \frac{N}{2}.
\end{align*}
\end{small}

\noindent Finally, we use $\Delta\mathcal{H}_{\mathrm{flow-var}}$ to indicate the information loss and leverage Stirling's formula to approach the factorial.

\begin{small}
\begin{align*}
    \Delta\mathcal{H}_{\mathrm{flow-var}} &= \mathcal{H}_{\mathrm{packet}} - \mathcal{H}_{\mathrm{flow-var}} \\
    &= N \ln K\sigma - \ln \frac{\sigma^2}{N} - \frac{N}{2} - \ln 2 (\frac{N}{2})! + \frac{N}{2}\ln\frac{N}{2} \\
    & (n!\approx \sqrt{2\pi n}(\frac{n}{e})^n) \\
    &= N \ln K\sigma - \ln \frac{\sigma^2}{N} - \frac{N}{2} + \frac{N}{2}\ln\frac{N}{2} - \ln 2\sqrt{\pi N}{(\frac{N}{2e})}^{(\frac{N}{2})} \\
    &= N \ln K\sigma - \ln \frac{\sqrt{4\pi N^3}}{\sigma^2}.
\end{align*}
\end{small}

\noindent Here, we complete the proof of the Theorem 4.

\begin{figure*}[t]
    \vspace{-2mm}
    \begin{center}
    \includegraphics[width=0.98\textwidth]{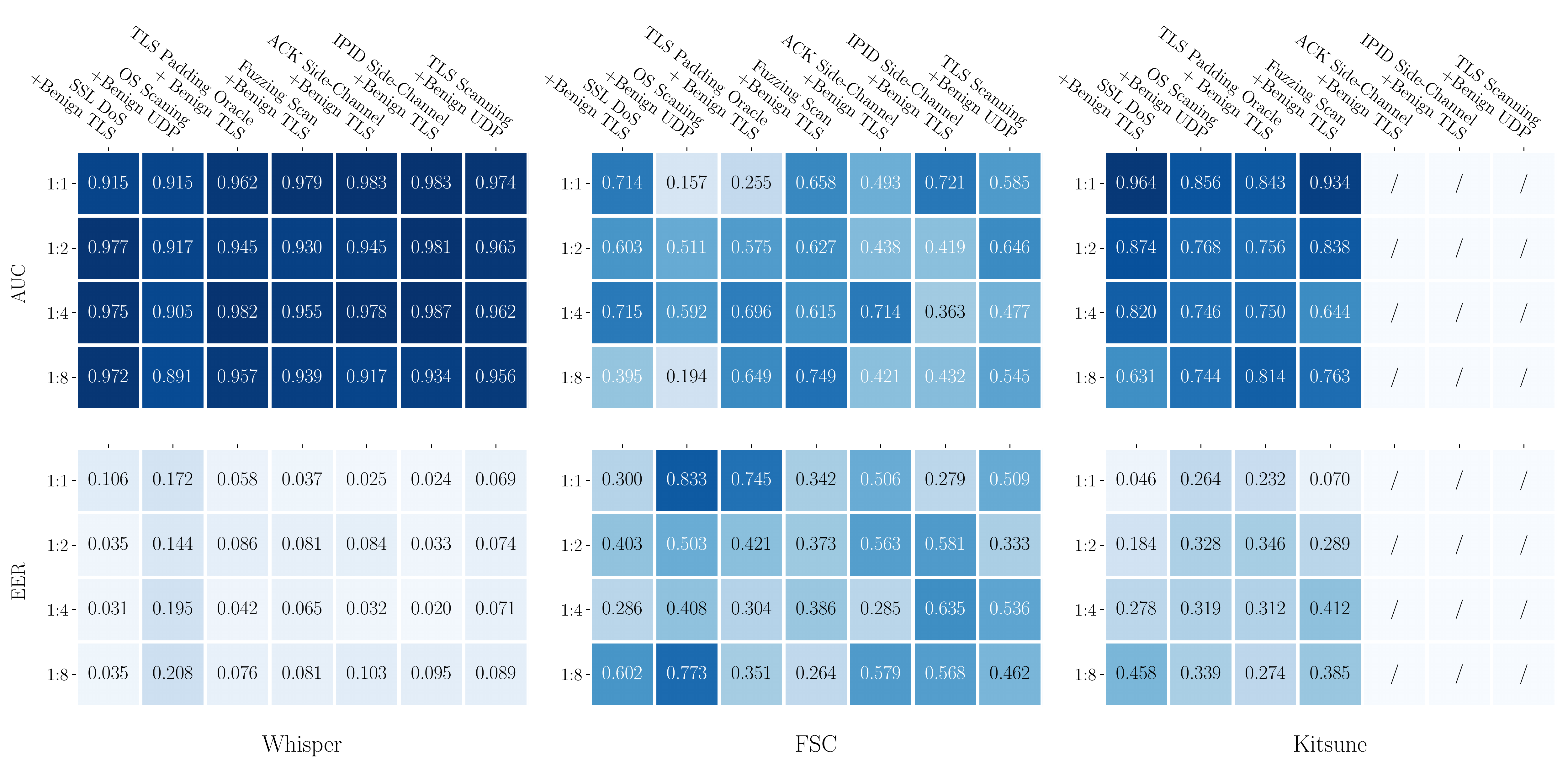}
    \vspace{-5mm}
 	\caption{Detection accuracy under 28 evasion attacks. During the attacks, in order to evade the detection, the attackers use different strategies to inject benign traffic.}
	\label{graph:robustness}
	\end{center}
\vspace{-2mm}
\end{figure*}

\section{Proof of Theorem 5 and Theorem 6} \label{section:appendix:5-6}
Without the loss of generality, we analyze $i^{th}$ kind of per-packet features, and denote its sampling sequence as $\vec{s}$. Based on the original assumption, we assume that Gaussian process $\mathcal{S}$ is strictly stationary with zero mean, i.e., $u(i)=0$ and $\sigma(i)=\sigma$. \NM\ extracts the frequency domain features of the per-packet feature sampling sequence $\vec{s}$ with the following steps:

\begin{enumerate}
    \item Perform linear transformation by multiplying $w_i$ on $\vec{s}$, for simplicity, we use $w$ to indicate $w_i$.
    \item Perform DFT on $w\vec{s}$. We denote the result as $\vec{F} = \mathscr{F}(w\vec{s})$ and its $i^{th}$ element as $\vec{F}_i = (a_i + jb_i)w$.
    \item Calculate modulus for the result of DFT. $\vec{P}$ denotes the result and $\vec{P}_i = (a_i^2 + b_i^2)w^2$ denotes its $i^{th}$ element.
    \item Perform logarithmic transformation on $\vec{P}$. $\vec{R}$ denotes the extracted frequency domain features for $\vec{s}$ and $\vec{R}_i = \ln(\vec{P}_i + 1) / C$ denotes its $i^{th}$ element.
\end{enumerate}

The property of Discrete Fourier Transformation: $\mathscr{F}(w\vec{s})=w\mathscr{F}(\vec{s})$, implies that:

\begin{align*}
    b_i =_{st} a_i, \quad a_i \sim \mathcal{N}(0, N\sigma^2).
\end{align*}

\noindent We estimate the overall differential entropy of the frequency domain features by ignoring the impact of the logarithmic transformation and obtain the entropy as $\mathcal{H}_{\mathrm{\NMM}}$. According to the properties of differential entropy and several inequalities about differential entropy, we obtain an estimation for the differential entropy of the frequency domain features:
\begin{small}
\begin{align*}
    \mathcal{H}_{\mathrm{\NMM}}  &= \mathcal{H}[\vec{P}] = \sum_{i=1}^{N} \mathcal{H}[P_i] \\
                        &= \sum_{i=1}^{N} \mathcal{H}[w^2(a_i^2 + b_i^2)]\\
                        &= N \ln w^2 + \sum_{i=1}^{N} \mathcal{H}[a_i^2 + b_i^2]\\
                        &= N \ln Nw^2 + \sum_{i=1}^{N} \mathcal{H}[(\frac{a_i}{\sqrt{N}})^2 + (\frac{b_i}{\sqrt{N}})^2],\\
                        & (t_i = (\frac{a_i}{\sqrt{N}})^2 + (\frac{b_i}{\sqrt{N}})^2, \quad t_i \sim \chi^2(2)),
\end{align*}
\begin{align*}
    \mathcal{H}_{\mathrm{\NMM}}  &= N \ln Nw^2 + \sum_{i=1}^{N} \mathcal{H}[t_i]\\
                                 &= N \ln Nw^2 + N (1 + \ln 2).
\end{align*}
\end{small}

\noindent We use $\Delta\mathcal{H}_{\mathrm{\NMM}}$ to indicate the information loss of \NM\ and get an estimation of the differential entropy loss of \NM:
\begin{small}
\begin{align*}
    \Delta\mathcal{H}_{\mathrm{\NMM}} &= \mathcal{H}_{\mathrm{packet}} - \mathcal{H}_{\mathrm{\NMM}} \\
                                     &= N\ln \frac{\sigma}{w^2} \sqrt{\frac{\pi}{2e}} - N \ln N.
\end{align*}
\end{small}

\noindent We complete the proof of Theorem 5. According to Theorem 1 - 4, we can obtain Theorem 6.

\newpage
\section{The Detailed Results of Robust Evaluation} \label{section:appendix:robust}
Figure~\ref{graph:robustness} shows the detailed detection results under different evasion attacks, i.e., seven types of malicious traffic mixed with benign traffic with four types of inject ratio. We observe that the injected benign traffic has negligible effects on the detection accuracy of \NM.

\begin{figure}[t]
    \begin{center}
    \includegraphics[width=0.49\textwidth]{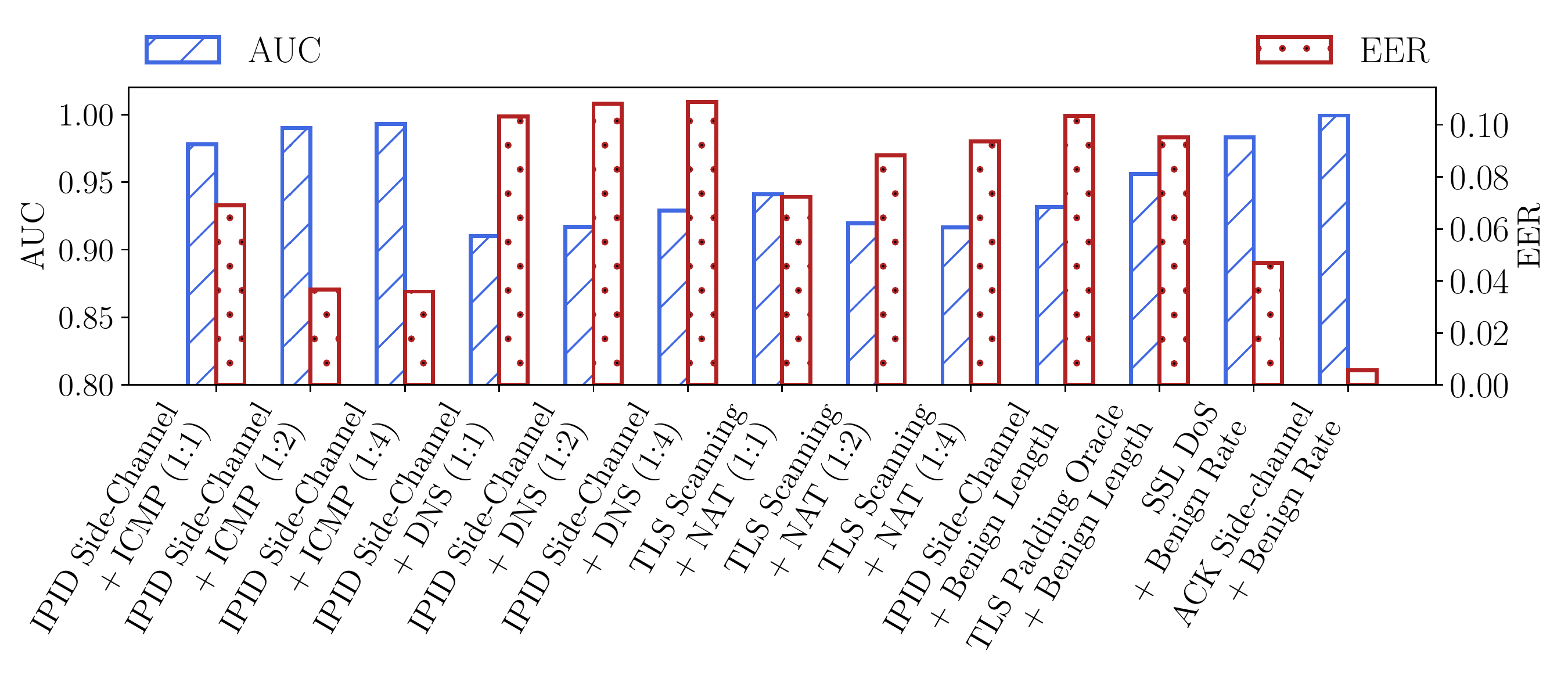}
    \vspace{-7mm}
 	\caption{Detection accuracy under sophisticated evasion strategies.}
	\label{graph:other}
	\end{center}
\vspace{-2mm}
\end{figure}

We also measure the effects of more sophisticated evasion strategies on the detection accuracy. The strategies include (i) injecting different types of benign traffic (i.e., ICMP, DNS, and outbound NAT traffic that includes various types of benign traffic), (ii) changing the rate of sending malicious packets according to the rate of benign TLS flows, (iii) manipulating the packet length in the malicious traffic according to the benign TLS packet length. Figure~\ref{graph:other} shows that the detection accuracy is not significantly impacted by the attacks, which is consistent with the results shown in Figure~\ref{graph:robustness}.

\end{document}